\documentclass[acmsmall]{acmart}

\usepackage{algorithmic}
\usepackage{graphicx}
\usepackage{flushend}
\usepackage{dblfloatfix}
\usepackage{textcomp}
\usepackage{xcolor}
\usepackage{url}
\usepackage{booktabs}
\usepackage{tcolorbox}
\usepackage{multirow}
\usepackage[colorinlistoftodos]{todonotes}
\usepackage{ulem}
\usepackage{balance}
\usepackage{colortbl}
\usepackage{listings}
\usepackage{enumitem}
\usepackage{comment}
\setlist[description]{leftmargin=\parindent,labelindent=\parindent}

\newlist{UP}{enumerate}{1}
\setlist[UP]{label=U\textsubscript{\arabic*}:}
\newlist{RP}{enumerate}{1}
\setlist[RP]{label=R\textsubscript{\arabic*}:}
\newlist{CP}{enumerate}{1}
\setlist[CP]{label=C\textsubscript{\arabic*}:}
\newlist{FP}{enumerate}{1}
\setlist[FP]{label=F\textsubscript{\arabic*}:}

\usepackage[percent]{overpic}%
\usepackage{tikz}
\usepackage{moresize}
\usepackage{caption}
\usepackage{subcaption}
\colorlet{punct}{red!60!black}
\definecolor{background}{HTML}{EEEEEE}
\definecolor{delim}{RGB}{20,105,176}
\colorlet{numb}{magenta!60!black}

\lstdefinelanguage{json}{
    basicstyle=\scriptsize,
    numberstyle=\scriptsize,
    stepnumber=1,
    numbersep=5pt,
    showstringspaces=false,
    breaklines=true,
    frame=lines,
    backgroundcolor=\color{background},
    literate=
     *{0}{{{\color{numb}0}}}{1}
      {1}{{{\color{numb}1}}}{1}
      {2}{{{\color{numb}2}}}{1}
      {3}{{{\color{numb}3}}}{1}
      {4}{{{\color{numb}4}}}{1}
      {5}{{{\color{numb}5}}}{1}
      {6}{{{\color{numb}6}}}{1}
      {7}{{{\color{numb}7}}}{1}
      {8}{{{\color{numb}8}}}{1}
      {9}{{{\color{numb}9}}}{1}
      {:}{{{\color{punct}{:}}}}{1}
      {,}{{{\color{punct}{,}}}}{1}
      {\{}{{{\color{delim}{\{}}}}{1}
      {\}}{{{\color{delim}{\}}}}}{1}
      {[}{{{\color{delim}{[}}}}{1}
      {]}{{{\color{delim}{]}}}}{1},
}

\definecolor{maroon}{cmyk}{0, 0.87, 0.68, 0.32}
\definecolor{halfgray}{gray}{0.55}
\definecolor{ipython_frame}{RGB}{207, 207, 207}
\definecolor{ipython_bg}{RGB}{247, 247, 247}
\definecolor{ipython_red}{RGB}{186, 33, 33}
\definecolor{ipython_green}{RGB}{0, 128, 0}
\definecolor{ipython_cyan}{RGB}{64, 128, 128}
\definecolor{ipython_purple}{RGB}{170, 34, 255}

\lstdefinelanguage{python}{
    morekeywords={access,and,break,class,continue,def,del,elif,else,except,exec,finally,for,from,global,if,import,in,is,lambda,not,or,pass,print,raise,return,try,while},
    morekeywords=[2]{abs,all,any,basestring,bin,bool,bytearray,callable,chr,classmethod,cmp,compile,complex,delattr,dict,dir,divmod,enumerate,eval,execfile,file,filter,float,format,frozenset,getattr,globals,hasattr,hash,help,hex,id,input,int,isinstance,issubclass,iter,len,list,locals,long,map,max,memoryview,min,next,object,oct,open,ord,pow,property,range,raw_input,reduce,reload,repr,reversed,round,set,setattr,slice,sorted,staticmethod,str,sum,super,tuple,type,unichr,unicode,vars,xrange,zip,apply,buffer,coerce,intern},
    sensitive=true,
    morecomment=[l]\#,
    morestring=[b]',
    morestring=[b]",
    morestring=[s]{'''}{'''},
    morestring=[s]{"""}{"""},
    morestring=[s]{r'}{'},
    morestring=[s]{r"}{"},
    morestring=[s]{r'''}{'''},
    morestring=[s]{r"""}{"""},
    morestring=[s]{u'}{'},
    morestring=[s]{u"}{"},
    morestring=[s]{u'''}{'''},
    morestring=[s]{u"""}{"""},
    literate=
    {á}{{\'a}}1 {é}{{\'e}}1 {í}{{\'i}}1 {ó}{{\'o}}1 {ú}{{\'u}}1
    {Á}{{\'A}}1 {É}{{\'E}}1 {Í}{{\'I}}1 {Ó}{{\'O}}1 {Ú}{{\'U}}1
    {à}{{\`a}}1 {è}{{\`e}}1 {ì}{{\`i}}1 {ò}{{\`o}}1 {ù}{{\`u}}1
    {À}{{\`A}}1 {È}{{\'E}}1 {Ì}{{\`I}}1 {Ò}{{\`O}}1 {Ù}{{\`U}}1
    {ä}{{\"a}}1 {ë}{{\"e}}1 {ï}{{\"i}}1 {ö}{{\"o}}1 {ü}{{\"u}}1
    {Ä}{{\"A}}1 {Ë}{{\"E}}1 {Ï}{{\"I}}1 {Ö}{{\"O}}1 {Ü}{{\"U}}1
    {â}{{\^a}}1 {ê}{{\^e}}1 {î}{{\^i}}1 {ô}{{\^o}}1 {û}{{\^u}}1
    {Â}{{\^A}}1 {Ê}{{\^E}}1 {Î}{{\^I}}1 {Ô}{{\^O}}1 {Û}{{\^U}}1
    {œ}{{\oe}}1 {Œ}{{\OE}}1 {æ}{{\ae}}1 {Æ}{{\AE}}1 {ß}{{\ss}}1
    {ç}{{\c c}}1 {Ç}{{\c C}}1 {ø}{{\o}}1 {å}{{\r a}}1 {Å}{{\r A}}1
    {€}{{\EUR}}1 {£}{{\pounds}}1
    {^}{{{\color{ipython_purple}\^{}}}}1
    {=}{{{\color{ipython_purple}=}}}1
    {+}{{{\color{ipython_purple}+}}}1
    {*}{{{\color{ipython_purple}$^\ast$}}}1
    {/}{{{\color{ipython_purple}/}}}1
    {+=}{{{+=}}}1
    {-=}{{{-=}}}1
    {*=}{{{$^\ast$=}}}1
    {/=}{{{/=}}}1,
    literate=
    *{-}{{{\color{ipython_purple}-}}}1
     {?}{{{\color{ipython_purple}?}}}1,
    identifierstyle=\color{black}\ttfamily,
    commentstyle=\color{ipython_cyan}\ttfamily,
    stringstyle=\color{ipython_red}\ttfamily,
    keepspaces=true,
    showspaces=false,
    showstringspaces=false,
    rulecolor=\color{ipython_frame},
    numberstyle=\tiny\color{halfgray},
    backgroundcolor=\color{ipython_bg},
    basicstyle=\scriptsize,
    keywordstyle=\color{ipython_green}\ttfamily,
}

\AtBeginDocument{%
  \providecommand\BibTeX{{%
    \normalfont B\kern-0.5em{\scshape i\kern-0.25em b}\kern-0.8em\TeX}}}

\setcopyright{acmcopyright}
\copyrightyear{2021}
\acmYear{2021}
\acmDOI{10.1145/1122445.1122456}

\acmJournal{TOSEM}
\acmVolume{XX}
\acmNumber{X}
\acmArticle{XXX}
\acmMonth{6}

\begin{document}

\newcommand\npmDE[1]{{GH-Node.js{#1}}}

\newcommand\bodin[1]{{\textcolor{red}{(Bodin: #1)}}}
\newcommand\raula[1]{{\textcolor{pink}{#1}}}
\newcommand\ishio[1]{{\textcolor{cyan}{#1}}}
\newcommand{\ma}[1]{\textit{\textcolor{blue}{(#1)}}}
\newcommand{\br}[1]{\textit{\textcolor{red}{(#1)}}}
\newcommand{\ct}[1]{\textit{\textcolor{purple}{(#1)}}}
\newcommand\todoV[2]{ {\colorbox{yellow}{\textcolor{red}{#2}}} {\todo[color=green!40]{\footnotesize{#1}}}}

\newcommand\RqZero{RQ1}
\newcommand\RqOne{RQ2}
\newcommand\RqTwo{RQ3}
\newcommand\RqZeroText{(\RqZero) \textit{Which features of an npm package do practitioners find relevant for assessing package quality?}}
\newcommand\RqOneText{(\RqOne) \textit{What features of an npm package correlate from different perspectives?}}
\newcommand\RqTwoText{(\RqTwo) \textit{What features of an npm package predict whether it is runnable or not?}}

\title{What makes a good Node.js package? Investigating Users, Contributors, and Runnability}

\author{Bodin Chinthanet}
\email{bodin.chinthanet.ay1@is.naist.jp}
\orcid{0000-0003-4439-1608}
\affiliation{%
  \institution{Nara Institute of Science and Technology}
  \city{Nara}
  \country{Japan}
}

\author{Brittany Reid}
\email{brittany.reid@adelaide.edu.au}
\orcid{0000-0001-7012-0655}
\author{Christoph Treude}
\email{christoph.treude@adelaide.edu.au}
\orcid{0000-0002-6919-2149}
\author{Markus Wagner}
\email{markus.wagner@adelaide.edu.au}
\orcid{0000-0002-3124-0061}
\affiliation{%
  \institution{University of Adelaide}
  \city{Adelaide}
  \country{Australia}
}

\author{Raula Gaikovina Kula}
\email{raula-k@is.naist.jp}
\orcid{0000-0003-2324-0608}
\author{Takashi Ishio}
\email{ishio@is.naist.jp}
\orcid{0000-0003-4106-699X}
\author{Kenichi Matsumoto}
\email{matumoto@is.naist.jp}
\orcid{0000-0002-7418-9323}
\affiliation{%
  \institution{Nara Institute of Science and Technology}
  \city{Nara}
  \country{Japan}
}

\renewcommand{\shortauthors}{B. Chinthanet et al.}

\begin{abstract}
The Node.js Package Manager (i.e., npm) archive repository serves as a critical part of the JavaScript community and helps support one of the largest developer ecosystems in the world.
However, as a developer, selecting an appropriate npm package to use or contribute to can be difficult.
To understand what features users and contributors consider important when searching for a good npm package, we conduct a survey asking Node.js developers to evaluate the importance of 30 features derived from existing work, including GitHub activity, software usability, and properties of the repository and documentation.
We identify that both user and contributor perspectives share similar views on which features they use to assess package quality.
We then extract the 30 features from 104,364 npm packages and analyse the correlations between them, including three software features that measure package ``runnability"; ability to install, build, and execute a unit test.
We identify which features are negatively correlated with runnability-related features and 
find that predicting the runnability of packages is viable.
Our study lays the groundwork for future work on understanding how users and contributors select appropriate npm packages.
\end{abstract}



\maketitle

\section{Introduction}
The Node.js Package Manager (i.e., npm) serves as a critical part of the JavaScript community and provides support to one of the largest developer ecosystems in the world, with over 1.7 million packages \cite{Web:npm}. 
These packages provide developers with useful features and libraries without the need to ``reinvent the wheel", with each package often depending on several others. Searching for a suitable third-party library is a well-known problem for software developers \citep{xia_what_2017}, especially in a massive network like the npm ecosystem.

Most research revolves around selection criteria based on quantifying a library goodness of fit for a certain scenario.
For example, works such as LibRec \citep{Thung:2013} and other proposed library recommendation techniques present a scenario where a developer is interested in using a library that has also been used in other similar projects \citep{Huang:2020, Nguyen:2020, Ouni:2017, Saied:2018, DelaMora:2018:which-lib, Pano:2018}. Other related work focuses on the motivations behind why a software developer selects a package. For example, \citet{Larios:2020} argue that software developers often choose libraries arbitrarily, without considering the consequences of their decisions.
These studies confirm that developers do struggle with library selection and updates \citep{Huang:2020, Wang:2018, Patra:2018, Zimmermann:2019, Derr:2017}. 
In all of these works, the common assumption is that the \textit{interested audience} of a library is a typical software developer who would like to adopt the package into their application.

Unlike related work, the novelty of this work is to provide a comprehensive investigation of features from previous work to quantify the goodness of a package by analysing the interested audience perspective. 
Based on the literature, we identified the audience from two overlapping perspectives:
\begin{enumerate}
    \item  \textit{Users} - who are looking to adopt a package into their applications. They are typically software developers \citep{Larios:2020, Web:choosing_lib}, who are interested in how well documented a package is and how it can be integrated into their existing project.
    \item  \textit{Contributors} - who are interested in contributing to a package. They are likely to be newcomer developers who view the package as a software project they would like to onboard \citep{Steinmacher:2014, Tan:2020}.
\end{enumerate}

To characterise Node.js packages published on npm registry (i.e., npm packages) from these two perspectives, we conducted an online survey to ask Node.js developers what features they consider before adopting or contributing to assessing the quality of npm packages.
We separate features of npm packages into four types: 
(1) \textit{Documentation}: the properties of the package documentation, 
(2) \textit{Repository}: the properties of the git repository, 
(3) \textit{Software}: the ability to install, build, and execute the test of the package, and
(4) \textit{GitHub activity}: the interaction between developers and the repository on GitHub.
We formulate the first research question as follows:
\begin{description}
    \item \RqZeroText
\end{description}
The key results of \RqZero~are that (1) users and contributors of npm packages share similar views on which features are important for selecting a good package; users focus more on how to use the package, while contributors focus on the contribution guideline and how to build and test the package,
(2) developers agree that software and documentation features are highly relevant to assess package quality, and
(3) developers believe that repository features do not belong to any perspectives and do not reflect the package quality.

\begin{table*}[h!b]
\centering
\caption{List of 30 features for package quality assessment presented in the survey. The features are grouped by their characteristic (feature type).}
\label{tab:features}
\scalebox{0.64}{
\begin{tabular}{@{}cll@{}}
\toprule
\textbf{Feature} & \multicolumn{1}{c}{\multirow{2}{*}{\textbf{Feature}}} & \multicolumn{1}{c}{\multirow{2}{*}{\textbf{Description}}} \\
\textbf{type} & & \\ \midrule
\multirow{10}{*}{\rotatebox{90}{Documentation}}
& [D1] hasAReadmemd & the existence of the README.md file in the root directory of the package \citep{Prana:EMSE2018, Ikeda:IEICE2019} \\
& [D2] linesInReadmemd & the number of lines in the README.md file \citep{Prana:EMSE2018, Ikeda:IEICE2019} \\
& [D3] numberOfCodeBlocks & the number of code snippets extracted from the README.md file \citep{Prana:EMSE2018, Ikeda:IEICE2019} \\
& [D4] readmemdCodeSnippetsNumber & the number of JavaScript snippets in the README.md file \citep{Ikeda:IEICE2019} \\
& [D5] hasAnInstallExample & the existence of an install example in the README.md file \citep{Prana:EMSE2018, Ikeda:IEICE2019, Hassan:ICSE_C2017} \\
& [D6] hasARunExample & the existence of a run example in the README.md file \citep{Prana:EMSE2018, Ikeda:IEICE2019} \\
& [D7] hasAContributionGuideline & the existence of a CONTRIBUTING.md file in the root directory of the package \citep{Prana:EMSE2018, Elazhary:ICSME2019, Kobayakawa:COMPSAC2017, Sholler:PLOS_CB2019} \\
& [D8] hasACodeOfConduct & the existence of a CODE\_OF\_CONDUCT.md file in the root directory of the package \citep{Prana:EMSE2018, Tourani:SANER2017} \\
& [D9] hasALicense & the package has either license file or license description in REAMDE.md file \citep{Prana:EMSE2018, Ikeda:IEICE2019, Almeida:ICPC2017, Munaiah:2017:Reaper} \\
& [D10] githubLink & the GitHub repository link of the package \citep{Ikeda:IEICE2019} \\ \midrule
\multirow{9}{*}{\rotatebox{90}{Repository}}
& [R1] hasASourceDirectory & the existence of a source directory (i.e., /src) within the root directory of the package \citep{Kalliamvakou:MSR2014} \\
& [R2] numberOfFiles & the number of files in the package \citep{Beller:MSR2017} \\
& [R3] numberOfJsFiles & the number of JavaScript files (.js) in the package \citep{Beller:MSR2017} \\
& [R4] numberOfHtmlFiles & the number of HTML files (.html) in the package \citep{Beller:MSR2017} \\
& [R5] sizeOfRepository & the size of the package repository in bytes \citep{Noei:EMSE2017} \\
& [R6] mostPopularFileExtension & the most popular file extension appearing in the repository \citep{Beller:MSR2017} \\
& [R7] hasATestDirectory & the existence of a test directory (i.e., /test, /tests) in the root directory of the package \citep{Munaiah:2017:Reaper} \\
& [R8] numberOfRepositoryTags & the number of git tags in the repository \citep{Joshi:MSR2019} \\
& [R9] numberOfCommits & the number of commits in the main branch of the package \citep{Beller:MSR2017} \\ \midrule
\multirow{3}{*}{\rotatebox{90}{Software}}
& [S1] ableToInstall & whether the package is able to install the project \citep{Macho:SANER2018} \\
& [S2] ableToBuild & whether the package is able to build (i.e., install dependencies, prepare working environment) \citep{Gallaba:ASE2018, Beller:MSR2017} \\
& [S3] ableToExecuteATest & whether the package is able to execute all test cases \citep{Beller:MSR2017} \\ \midrule
\multirow{8}{*}{\rotatebox{90}{GitHub activity}}
& [G1] numberOfNewcomerLabels & the number of newcomer labels in the GitHub issues \citep{Tan:2020} \\
& [G2] numberOfContributors & the number of GitHub contributors in the package \citep{Munaiah:2017:Reaper, Gousio:2013} \\
& [G3] numberOfIssues & the number of GitHub issues \citep{Gousio:2013, Zhou:TSE2020} \\
& [G4] numberOfPullRequests & the number of GitHub pull requests \citep{Gousio:2013} \\
& [G5] mostRecentIssue & the most recent GitHub issue date in an epoch format \citep{Gousio:2013} \\
& [G6] mostRecentPullRequest & the most recent pull request date in an epoch format \citep{Gousio:2013} \\
& [G7] latestRepositoryUpdate & the most recent committed date in an epoch format \citep{Gousio:2013} \\
& [G8] latestReadmemdUpdate & the most recent README.md update date in an epoch format \citep{Gousio:2013} \\
\bottomrule
\end{tabular}
}
\end{table*}

To investigate what trade-offs exist between features, we collected diverse data related to Node.js packages to create \npmDE, a dataset of 723,218 Node.js package repositories.
First, we extracted and analysed the npm package features from 104,364 npm packages to understand the correlation among the features.
We then explored the possibility of predicting package runnability since software features are highly relevant for assessing package quality, as shown in our survey.
Through the two lenses of perspectives and extracted features, we formulate the two following research questions to guide our study: 
\begin{description}
    \item \RqOneText
\end{description}
The key results of \RqOne~are that (1) features from the same type usually have strong correlations.
(2) Software features are negatively correlated with other features.
(3) Features that are less likely to be considered by any perspective, such as a number of files, are correlated with the ability to build a package.
\begin{description}
    \item \RqTwoText
\end{description}
The key results of \RqTwo~are that (1) predicting runnability of the package is viable.
(2) Based on the permutation feature importance, we find that repository features are important for predicting the runnability.

In summary, this paper presents the following contributions:
\begin{itemize}
    \item A definition of the `goodness' of an npm package from two perspectives, user and contributor. 
    \item A survey of Node.js developers to investigate how different perspectives select npm packages.
    \item Three measures of the runnability of an npm package.
    \item \npmDE, a dataset that contains a curated set of 723,218 Node.js packages containing repository and social interaction information.
    \item A large scale analysis of 104,364 npm packages for correlations and predictions of runnability. 
\end{itemize}

\section{Audience Perspectives of an npm package}
\label{sec:perspective}
In this paper, we characterise the audience of an npm package into two different perspectives. These perspectives are based on the various ways in which npm packages or software projects, in general, have been discussed in related work:

\begin{enumerate}
    \item \textit{User perspective} - according to \citet{lethbridge2003software}, apart from very good code design and reusability aspects, high quality documentation is the key to learning a software system. 
    \citet{Web:choosing_lib} explained that ``The greatest library in the world would fail if the only way to learn it was reading the code (and, in fact, it already has to a large extent). Some packages have managed to overcome this by way of lots of unofficial documentation -- blog entries and the like -- but there is absolutely no substitute for full, well-written documentation." 
    Therefore, we consider the user perspective essential to our investigation of what makes a good Node.js package.
    
    \item  \textit{Contributor perspective} - according to \citet{Steinmacher:2014}, newcomers can be attracted to a project with characteristics such as license type, project visibility, or a number of existing contributors. Moreover, providing good support for the onboarding of newcomers to contribute to a project is important. It can be done by identifying good first issues and creating informative descriptions for them \citep{Tan:2020}. A package that is good for users (e.g., easy to install) might not necessarily be good for contributors (e.g., presence of a CONTRIBUTING.md) and vice versa. Investigating such different perspectives is one of the goals of this work.
    
\end{enumerate}

\subsection{Developer Survey on Perspective}

To answer \RqZeroText~we start our investigation with a list of 30 package features derived from related work.
We extracted these features from papers on npm in particular or software reuse in general.
Table~\ref{tab:features} shows the 30 features we identified along with the reference(s) for each.
We group the features into four types based on their characteristics: (1) \textit{Documentation}, (2) \textit{Repository}, (3) \textit{Software}, and (4) \textit{GitHub activity}.
However, related work has not validated the usefulness of these features to contributors and users.
To address this, we conducted an online survey to identify developer opinions on what perspectives npm features belong to and what features are relevant for assessing npm package quality.

In the survey, we first asked the developers ``which perspective do these features belong to?" for each of the 30 features, to which they can answer either (1) user perspective, (2) contributor perspective, (3) both of them, or (4) none of them.
To summarise the result of this question, we assign the top voted choice as the perspective of each feature.
In the case of the top voted choices are ``user" and ``contributor" (i.e., scores are tied), we assign ``both of them" to the feature.
We also confirmed that there were no cases where the ``none of them" choice tied with others.
After that, we asked the developers about their demographics, i.e., their experience with Node.js and npm packages, including their contributions on GitHub.
Next, we asked developers to evaluate ``how relevant are these features for assessing package quality?" for each of the 30 features using a five-point Likert scale, i.e., ranging from strongly agree to strongly disagree with a neutral option the middle strongly.
To find participants for the survey, we contacted 2,150 npm developers and received 33 responses.

\begin{figure}[t]
\centering%
\includegraphics[width=0.5\textwidth]{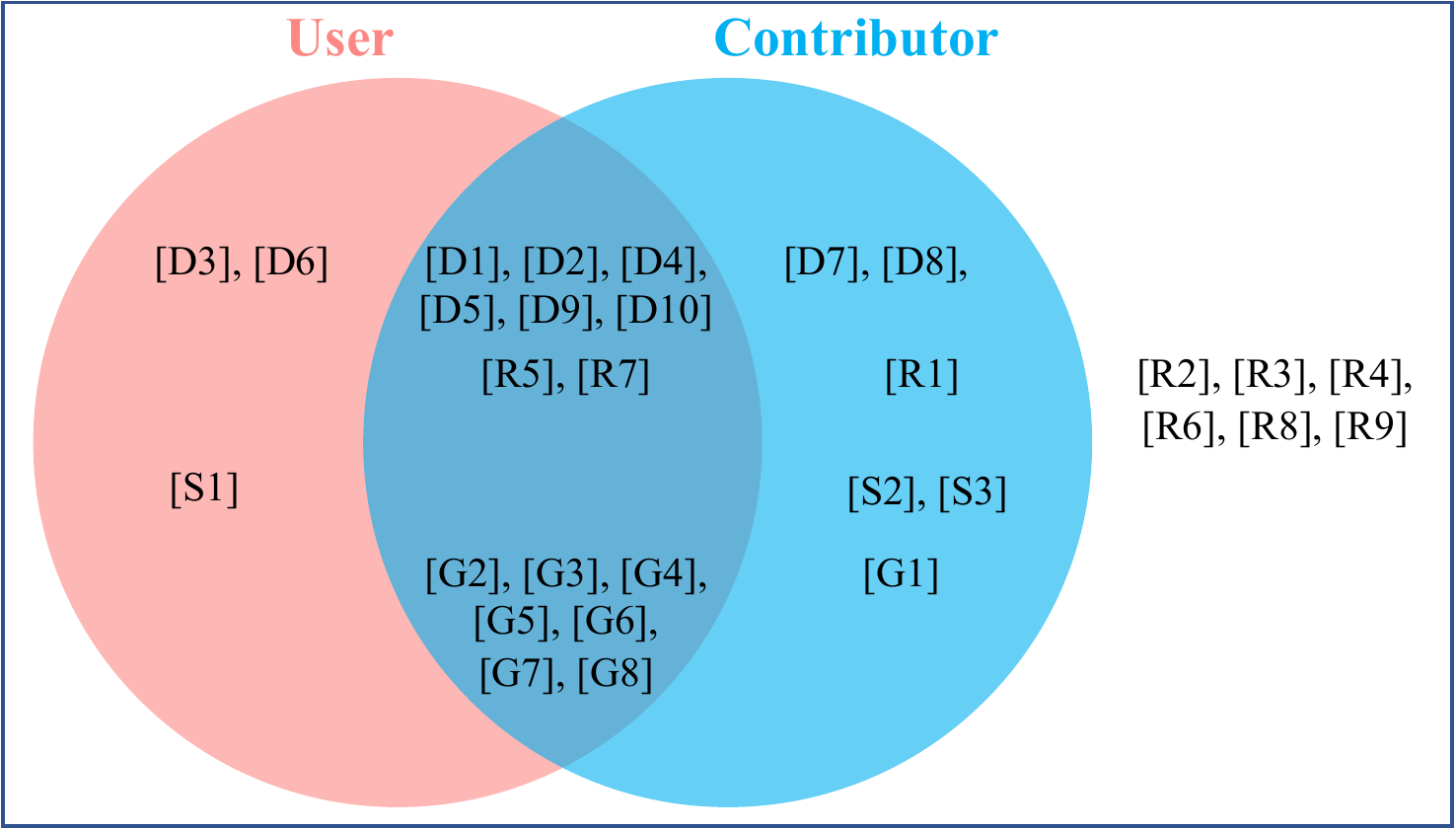}
\caption{Respondents mapping of features to the perspectives.}%
\label{fig:perspective}%
\end{figure}

\begin{table}[]
\centering
\caption{Demographic of the respondents (33 Node.js developers).}
\label{tab:survey_demographic}
\begin{tabular}{@{}lr@{}}
\toprule
\multicolumn{2}{c}{\textbf{How many years of experience do you have with Node.js?}} \\ \midrule
2 to 3 years & 3 \\
4 to 5 years & 11 \\
6 to 7 years & 12 \\
8 to 9 years & 3 \\
10 years or more & 4 \\ \midrule
\multicolumn{2}{c}{\textbf{What do you identify yourself as?}} \\ \midrule
User & 19 \\
Contributor & 2 \\
Both & 12 \\ \midrule
\multicolumn{2}{c}{\textbf{How often do you use npm packages in your projects?}} \\ \midrule
Once a month or more often & 30 \\
Less than once a month but more than once per year & 2 \\
Less than once per year & 1 \\ \midrule
\multicolumn{2}{c}{\textbf{How often do you contribute to npm packages on GitHub?}} \\ \midrule
Once a month or more often & 8 \\
Less than once a month but more than once per year & 18 \\
Less than once per year & 6 \\
Never & 1 \\ \bottomrule
\end{tabular}
\end{table}

Table \ref{tab:survey_demographic} shows the demographics of the participants.
All have at least two years of experience with Node.js and npm. 19 developers described themselves as only a user of packages, two as a contributor to packages and 12 as both a user and contributor.
Regarding the usage of npm packages, 30 developers responded that they used packages once a month or more often, two used them less than once a month but more than once per year, and the only one used npm packages less than once per year.
Eight developers contribute to packages on GitHub once a month or more often, 18 less than once a month but more than once per year, six contribute less than once per year, and only one never makes any contributions.

Figure \ref{fig:perspective} shows which perspective package features belong to according to the opinion of developers.
We find that half of the features are shared among both user and contributor perspectives.
If features are specific to either perspective, users focus on the example of code or snippet (D3, D6) and ability to install the package (S1).
On the other hand, contributors focus on guideline and code of conduct (D7, D8), source code directory (R1), ability to build and test (S2, S3), and newcomer labels in GitHub issues (G1).
Interestingly, most repository features (R2, R3, R4, R6, R8, R9) do not belong to any perspectives since their usefulness does not convince participants to assess the package quality.

\begin{figure*}
    \centering
    \begin{subfigure}[h]{\textwidth}
        \centering
        \caption{Documentation features (Median agreement: 73\%, neutral: 24\%, disagreement: 6\%)}
        \includegraphics[width=0.7\textwidth]{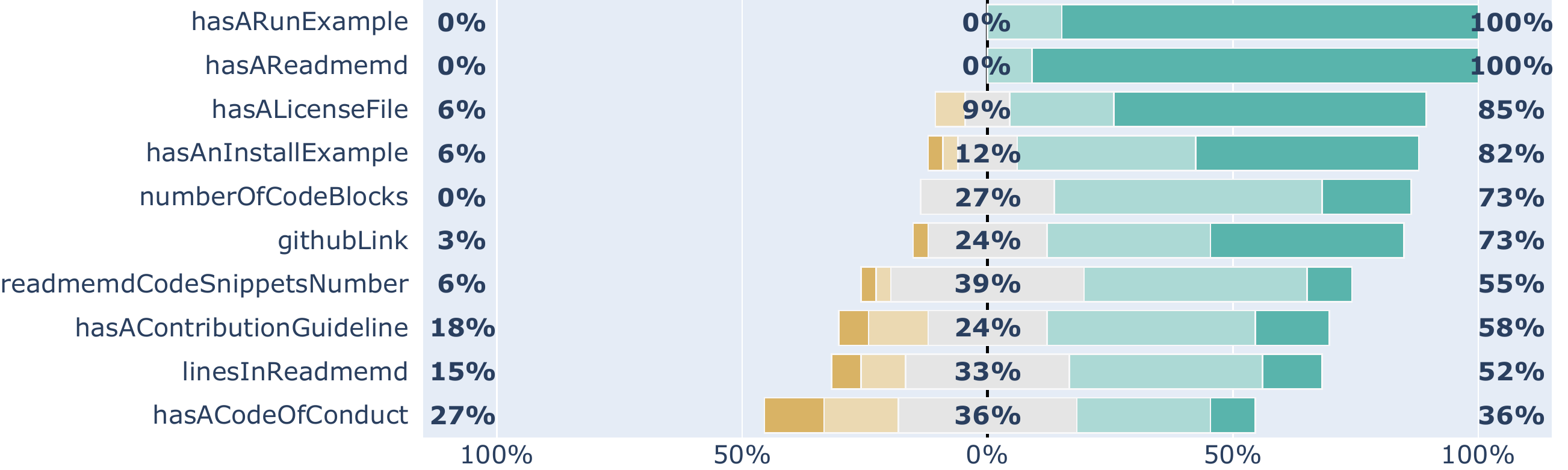}
        \label{subfig:likert_documentation}
    \end{subfigure}
    \begin{subfigure}[h]{\textwidth}
        \centering
        \vspace{0.5em}
        \caption{Repository features (Median agreement: 27\%, neutral: 34.5\%, disagreement: 42\%)}
        \includegraphics[width=0.7\textwidth]{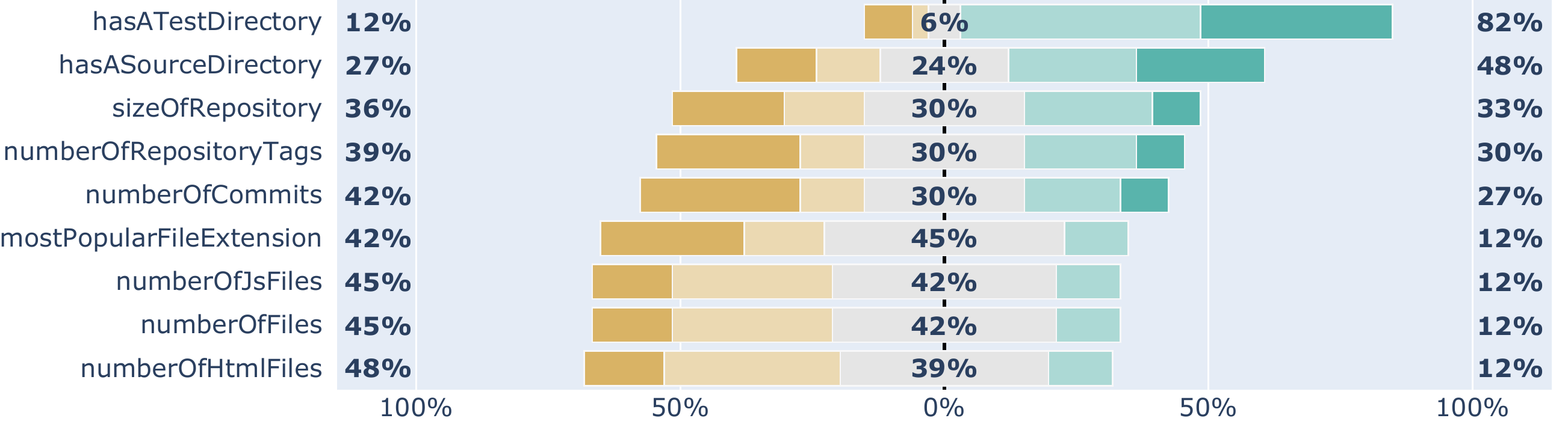}
        \label{subfig:likert_repository}
    \end{subfigure}
    \begin{subfigure}[h]{\textwidth}
        \centering
        \vspace{0.5em}
        \caption{Software features (Median agreement: 79\%, neutral: 15\%, disagreement: 6\%)}
        \includegraphics[width=0.7\textwidth]{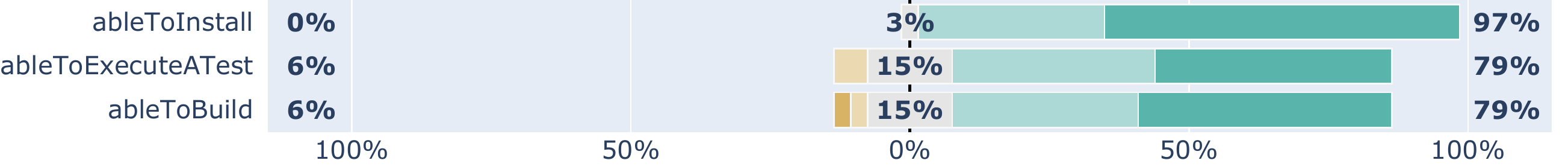}
        \label{subfig:likert_software}
    \end{subfigure}
    \vspace{1em}
    \begin{subfigure}[h]{\textwidth}
        \centering
        \vspace{0.5em}
        \caption{GitHub activity features (Median agreement: 43.5\%, neutral: 37.5\%, disagreement: 19.5\%)}
        \includegraphics[width=0.7\textwidth]{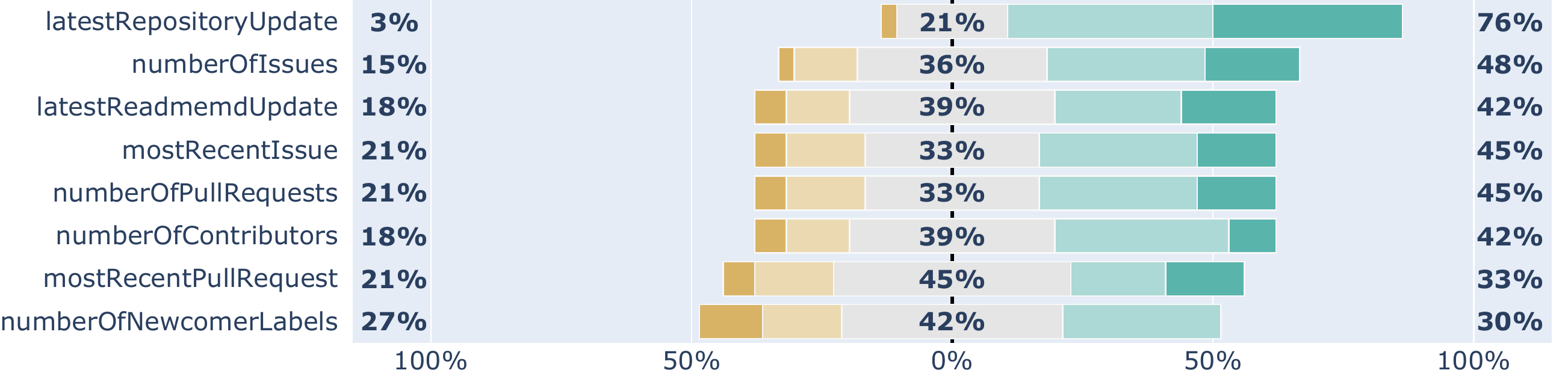}
        \label{subfig:likert_github_activity}
    \end{subfigure}
    \caption{Survey results on the relevance of features for package quality assessment. Left hand (yellow) shows levels of disagreement, middle (grey) shows neutral, and right (green) shows levels of agreement.}
    \label{fig:likert}
\end{figure*}

Figure \ref{fig:likert} shows the survey results on how relevant the features are for assessing package quality.
The green, grey, and yellow bars represent the level of agreement, neutrality, and disagreement from developer votes.
Overall, we find that developers agree that documentation, GitHub activity, and software features could be used to measure the quality of npm packages -- on average, 73\%, 43.5\%, and 79\% respectively agree (i.e., agree and strongly agree).
There are three features that get more than 90\% of agreement includes (1) hasARunExample (100\%), (2) hasAReadmemd (100\%), and (3) ableToInstall (97\%).
On the other hand, developers disagree that repository features can reflect the quality of packages -- on average, 42\% of the respondents disagree (i.e., disagree and strongly disagree). 

\begin{tcolorbox}
    \textbf{Summary:} 
    Assessing a package quality from both perspectives is slightly different.
    Users focus on how to use the package, while contributors focus on the contribution guidelines and how to build and test the package.
    Developers agree that software and documentation features are highly relevant for package quality assessment; in contrast, repository features are not. 
\end{tcolorbox}

\section{\npmDE: A Node.js Repository and Interaction Dataset}
To extract the required features needed by the two perspectives, we compose our own dataset of npm repositories.
\npmDE~is an open dataset contains a curated snapshot of Node.js and npm packages that focuses on the social, technical, and documentation aspects.
\npmDE~is based on two existing datasets (GHTorrent and Libraries.io).
GHTorrent provides a mirror of git repositories and developer interactions gathered from GitHub \citep{Gousio:2013}, while
Libraries.io provides meta-data and relationships among packages hosted on popular software ecosystems, e.g., npm, Maven, and PyPI \citep{Web:libraries.io}.

Table \ref{tab:data_collection} shows the summary statistics of \npmDE~from November 30, 2020, with 723,218 Node.js repositories, which only 104,364 repositories are identified as npm packages. 
We also collected 1,960,345 issues, 1,083,828 pull requests, and 283,360 contributors associated with npm packages by using the GitHub API.
In total, we took four months, from August to December 2020, to acquire and process all information for \npmDE.
The dataset is available at this following link: \url{https://doi.org/10.5281/zenodo.5010160}.
\begin{table}[t]
\centering
\caption{Dataset Snapshot Statistics. The full dataset estimations are approximate values.}
\label{tab:data_collection}
\begin{tabular}{@{}l|c@{}}
\toprule 
\multicolumn{2}{c}{\textbf{Node.js and npm Package Repository Information}} \\ \midrule
 Repository Snapshot & Nov 2, 2020 \\
 \# Node.js packages & 723,218 \\
 \# total npm packages  & 104,364 \\
 \hspace{1em}- \# last update 2013 & 7 \\
 \hspace{1em}- \# last update 2014 & 9,867 \\
 \hspace{1em}- \# last update 2015 & 25,680 \\
 \hspace{1em}- \# last update 2016 & 26,115 \\
 \hspace{1em}- \# last update 2017 & 15,057 \\
 \hspace{1em}- \# last update 2018 & 8,869 \\
 \hspace{1em}- \# last update 2019 & 7,642 \\
 \hspace{1em}- \# last update 2020 & 11,127 \\
 \hspace{1em}- \# commits & 6,402,982 \\
 \hspace{1em}- \# repository tags & 674,258 \\
 \bottomrule
\end{tabular}
\end{table}

\begin{table}[h]
\centering
\caption{Dataset for our experiment.}
\label{tab:experiment_data}
\begin{tabular}{@{}l|r@{}}
\toprule 
\multicolumn{2}{c}{\textbf{Runnable Code}} \\ \midrule
\# repositories (last updated in 2020) & 11,127 \\
\# successfully built & 8,199  \\
\hspace{1em} - \# passed tests  & 607 \\
\hspace{1em} - \# failed tests & 4,262 \\
\hspace{1em} - \# no test & 3,330 \\
\# unsuccessfully built & 2,928 \\ \midrule
\multicolumn{2}{c}{\textbf{Executable Documentation}} \\ \midrule
\# repositories & 104,364 \\ 
\# collected code snippets & 233,826 \\ 
\hspace{1em} - Max & 302 \\
\hspace{1em} - Min & 0 \\
\hspace{1em} - Mean & 2.2405 \\
\hspace{1em} - Median & 1.0000 \\
\hspace{1em} - SD & 4.8990 \\
\# successfully installed & 97,006 \\
\# executed code snippets & 220,324 \\
\hspace{1em} - \# successfully executed & 33,484 \\
\hspace{1em} - \# unsuccessfully executed & 186,840 \\
\bottomrule
\end{tabular}
\end{table}

\section{Experiment Setup}
To answer \RqOne~and \RqTwo, we extracted a subset of \npmDE~as shown in Table~\ref{tab:experiment_data}. 
For our data preparation, we first extract metrics that relate to each perspective.
Note that to answer \RqOne, we focus on all 11,127 packages that were last updated in 2020 from the total of 104,364 packages.
Our motivation was to retrieve projects that were likely to be active.

\subsection{Runnable Code: Build and Run Tests}
\label{subsec:build_test}
Building a package (79\%) and running its unit tests (also 79\%) are two features that received positive feedback in \RqZero~survey. 
They are also among the first tasks that contributors do to ensure that the selected package is ready to run and test new features or fixes in the local environment. 
To confirm that an npm package is runnable from its source code, we automatically create a virtual environment, build, and test the package from scratch.

Our approach to building and testing each package from its repository consists of four steps:
\begin{enumerate}
    \item Create a docker container to construct an isolated environment.
    \item Clone the repository of the npm package to the docker container.
    \item Build the repository by using \lstinline{npm install --no-audit}.
    \item If the package is built successfully, then execute \lstinline{npm test} to run a unit test script as detailed in the package.json file.
\end{enumerate}
From our trial and error, we set a timeout of five minutes for processing to detect frozen processes.

As shown in Table \ref{tab:experiment_data}, we selected 11,127 repositories with at least one commit in 2020 as the input of the runnable code experiment.
We find that 8,199 packages (73.68\%) are successfully built.
From these built packages, only 607 packages (7.40\%) have at least one test case and are able to pass all their test cases.
The rest of successful built packages are failed to pass any test cases (51.98\%) or no test case available (40.61\%).
Note that the building and the testing process took around 21 days of execution.

\subsection{Runnable Package: Install and Execute Code Snippets}
Installing a package (97\%) and trying the run example (100\%) are two features from the user perspective, which also received a positive result from \RqZero.
Both features are among the first tasks that package users do to ensure that the selected package is executable in their applications.
To confirm that an npm package is runnable after installed in any application, we automatically create a virtual environment, install, and execute the extracted code snippets.

We extracted 233,826 Node.js code snippets from the README files of all 104,364 repositories (last updated from 2013 up to 2020).
Example code in markdown files can be identified by the surrounding special characters (\lstinline{```}) that allow for the rendering of code blocks, and additional language information may be provided to allow for syntax highlighting; for example, adding the tag \texttt{js} to the same line (\lstinline{```js}) will enable JavaScript syntax highlighting on GitHub.
We extracted these code blocks, along with any language data, and then filtered our dataset to contain only Node.js code.
To do this, we discarded any code blocks that were not in the JavaScript language, such as bash commands used to demonstrate package installation, but kept those without a specified language, as not all READMEs utilise syntax highlighting. 
Then, to discard more unrelated code snippets, we filtered out common install commands such as \texttt{npm install}. 
Within Node.js repositories, it is also common to see code blocks containing the results of the previous snippet, often in the form of a JSON data object or array; as such, we filtered out any singular objects or arrays from our set of snippets.

To simulate how developers include the package in their projects and test its usage, we use a similar approach to Section~\ref{subsec:build_test}.
Instead of building the package itself, we built the empty Node.js package, which depends on the selected npm package.
After that, we executed the code snippets inside our empty package.

As shown in Table \ref{tab:experiment_data}, we selected all 104,364 npm package repositories as the input of the executable documentation experiment.
We found that 97,006 packages (92.95\%) are successfully installed.
From these installed packages, we found that only 64,280 packages (61.59\%) have at least one code snippet in their README.
In the end, we executed 220,324 code snippets and found that only 33,484 (15.20\%) of them were successfully executed without any error.
Note that the code snippet executing process took around 14 days of execution.

\begin{figure*}[t]
\centering%
\includegraphics[height=49mm,trim={10 216.5 0 0},clip]{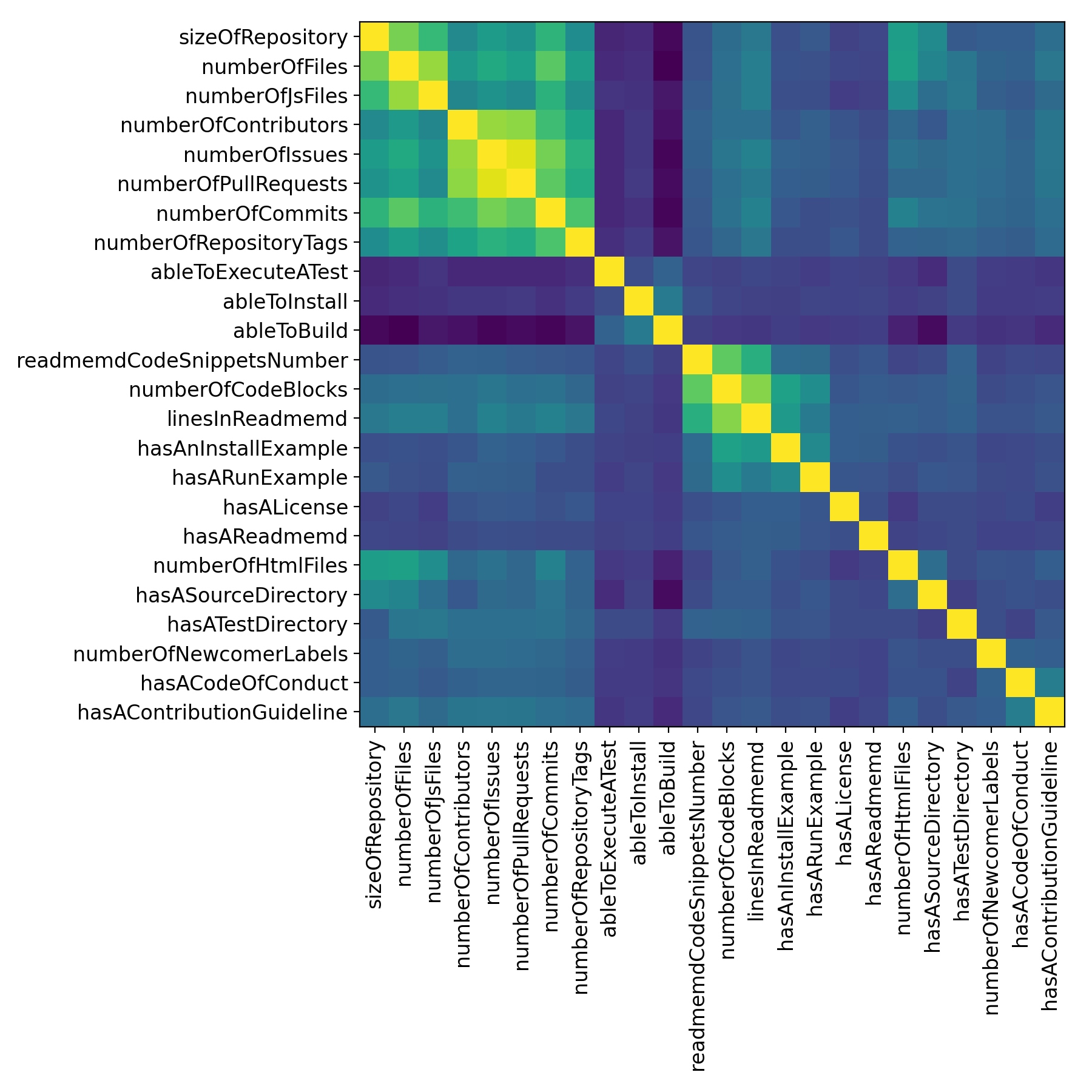}%
\hspace{2mm}%
\includegraphics[height=48mm,trim={0 0 0 0},clip,angle=0,origin=c]{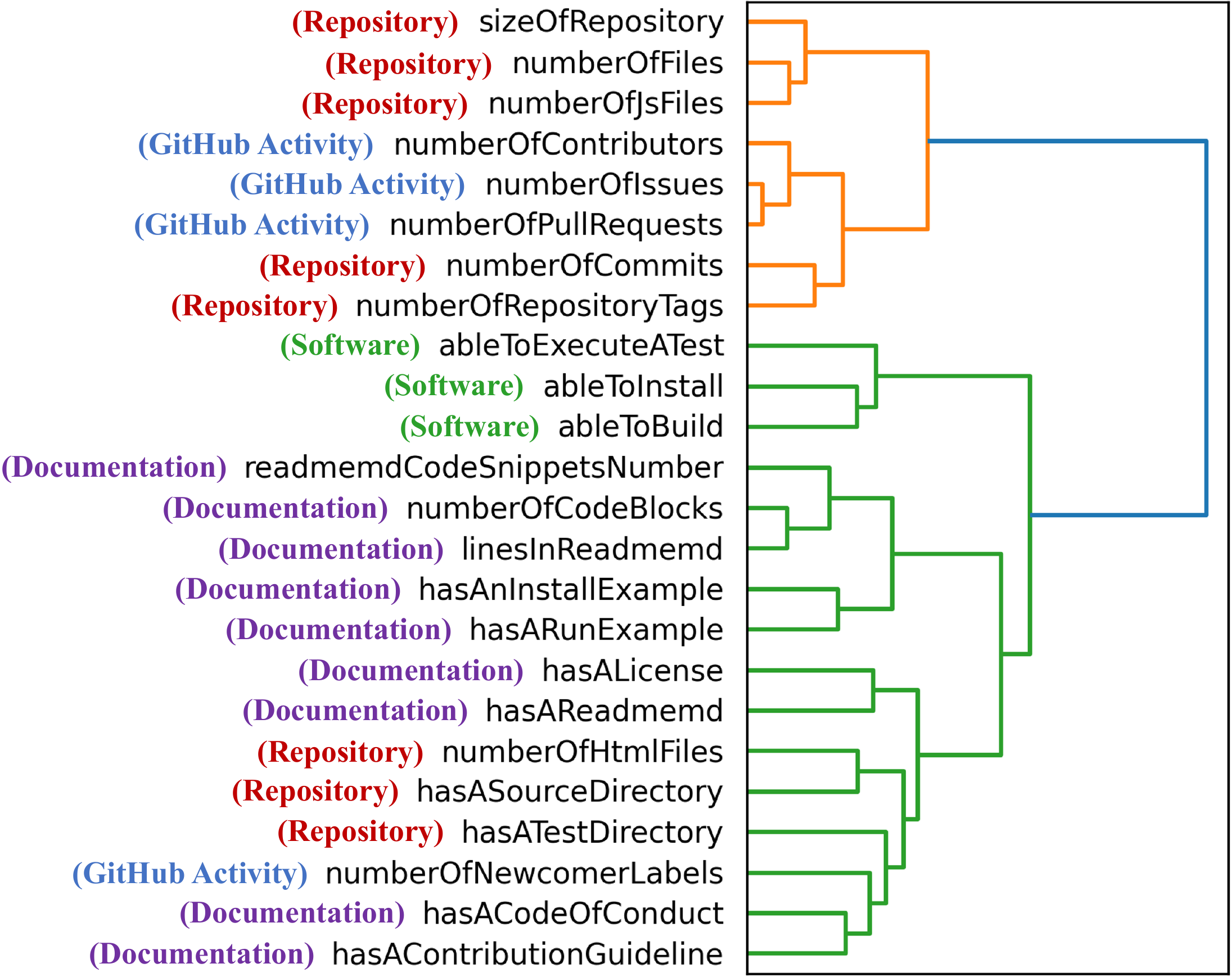}%
\caption{Features: correlations (left) and clustering (right). Lighter fields correspond to a strong positive correlation between the features, and darker fields to a strong negative correlation. X-labels are omitted as they follow the order (optimised to co-locate correlated features) of the y-labels. The dendrogram groups correlated features closely together. Shown are the correlations based on the 11,127 data points for which all feature values are available; not shown are the four timestamp-related features.}%
\label{fig:repos:instanceclusters}%
\end{figure*}

\newcommand{\plotreal}[2]{\begin{minipage}{30mm}\centering\ssmall{\textsf{#1}}\\
\includegraphics[width=30mm,trim={0 0 0 262},clip]{images/#2}%
\end{minipage}}%

\newcommand{\plotbinary}[2]{\begin{minipage}{30mm}\centering\ssmall{\textsf{#1}}\\
\includegraphics[width=30mm,trim={0 262 0 0},clip]{images/#2}%
\end{minipage}}%

\begin{figure*}[]\centering
\rotatebox{90}{\hspace{-36mm}\textsf{\scriptsize non-binary features\vspace{2mm}}} \rotatebox{90}{\hspace{-59mm}\rule{68mm}{0.4pt}}\hspace{3mm}
\plotreal{latestReadmemdUpdate}{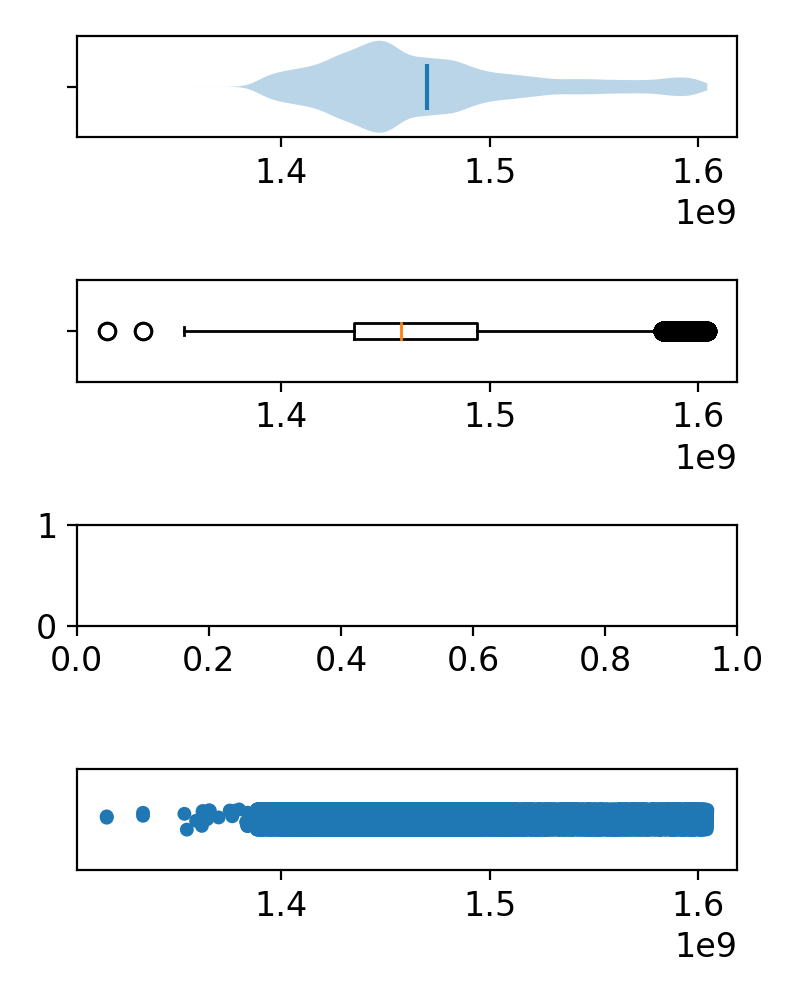}
\plotreal{latestRepositoryUpdate}{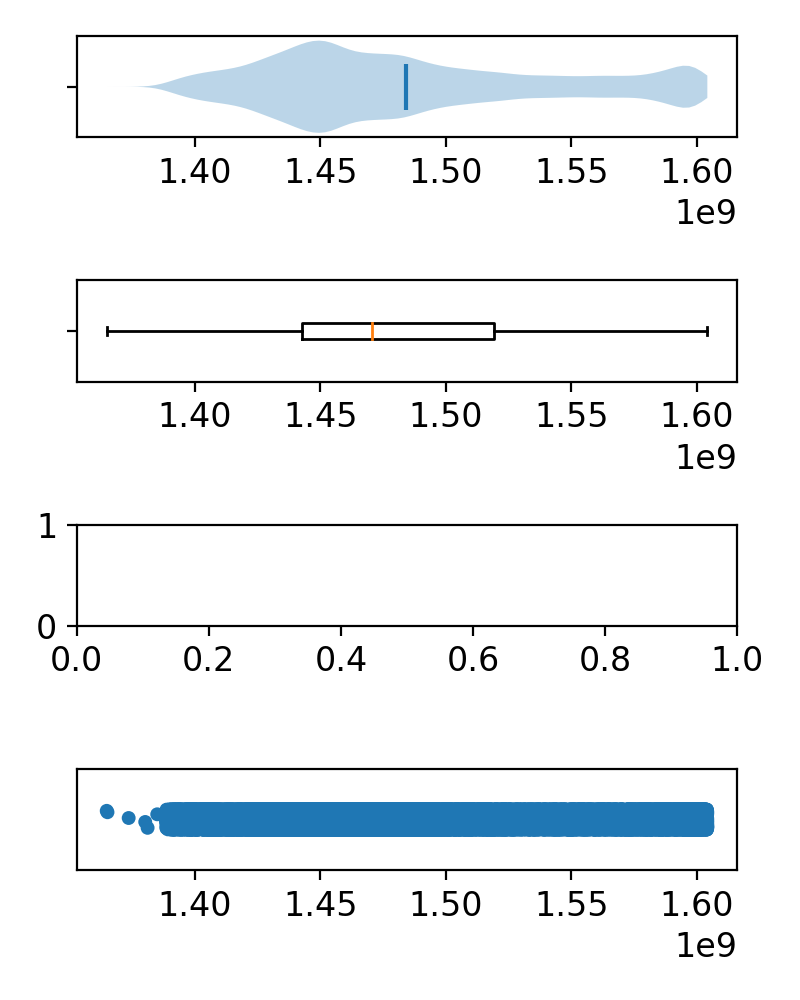}
\plotreal{linesInReadmemd}{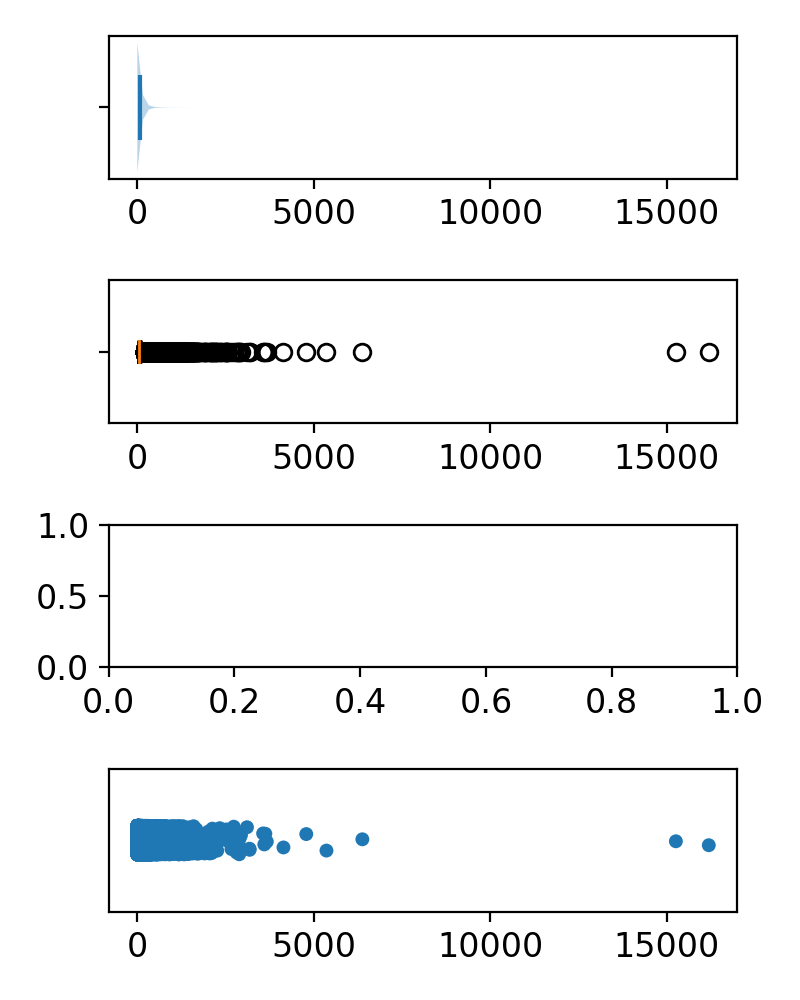}
\plotreal{mostRecentIssue}{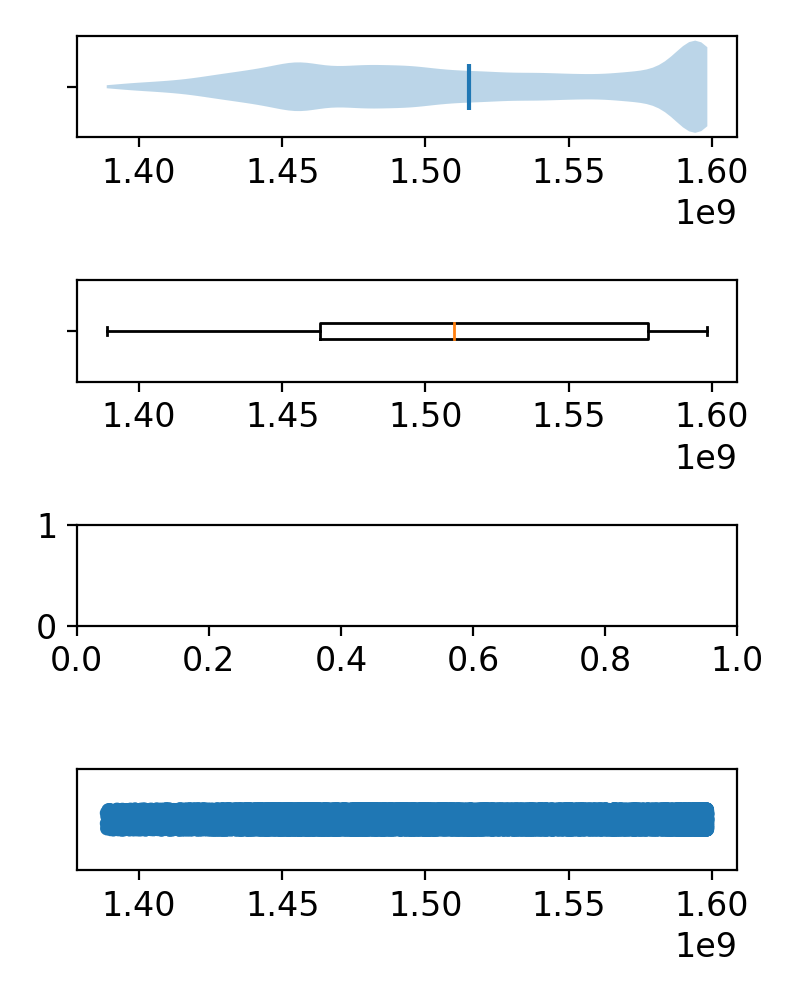}\\
\hspace{8mm}\plotreal{mostRecentPullRequest}{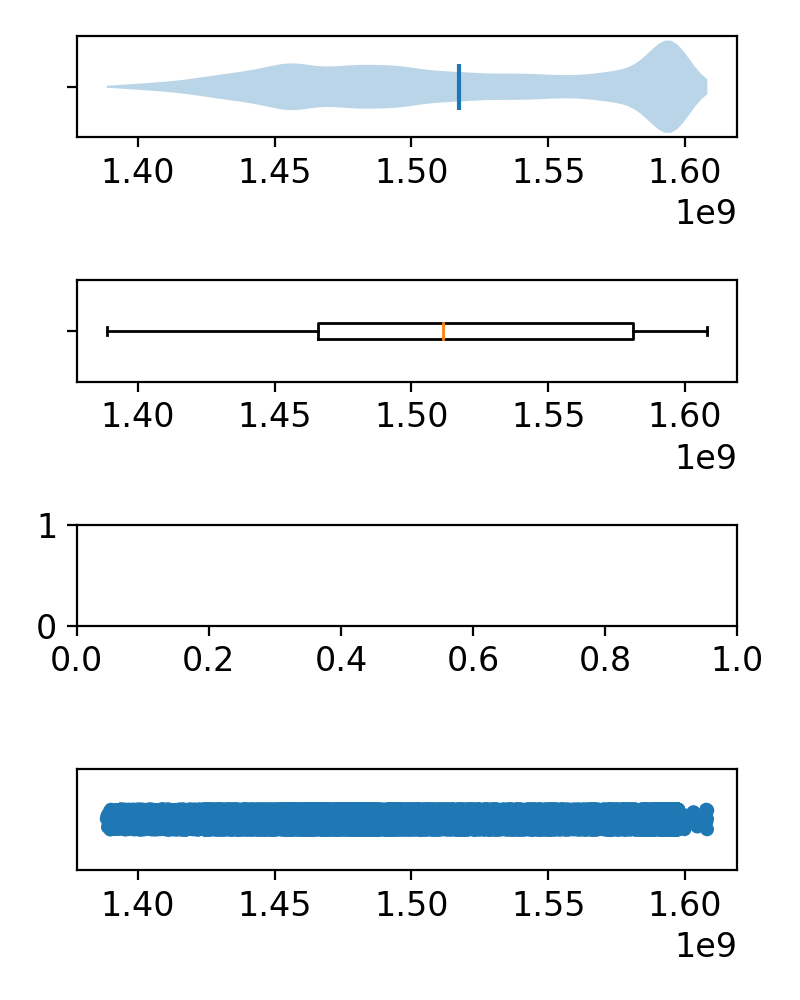}
\plotreal{numberOfCodeBlocks}{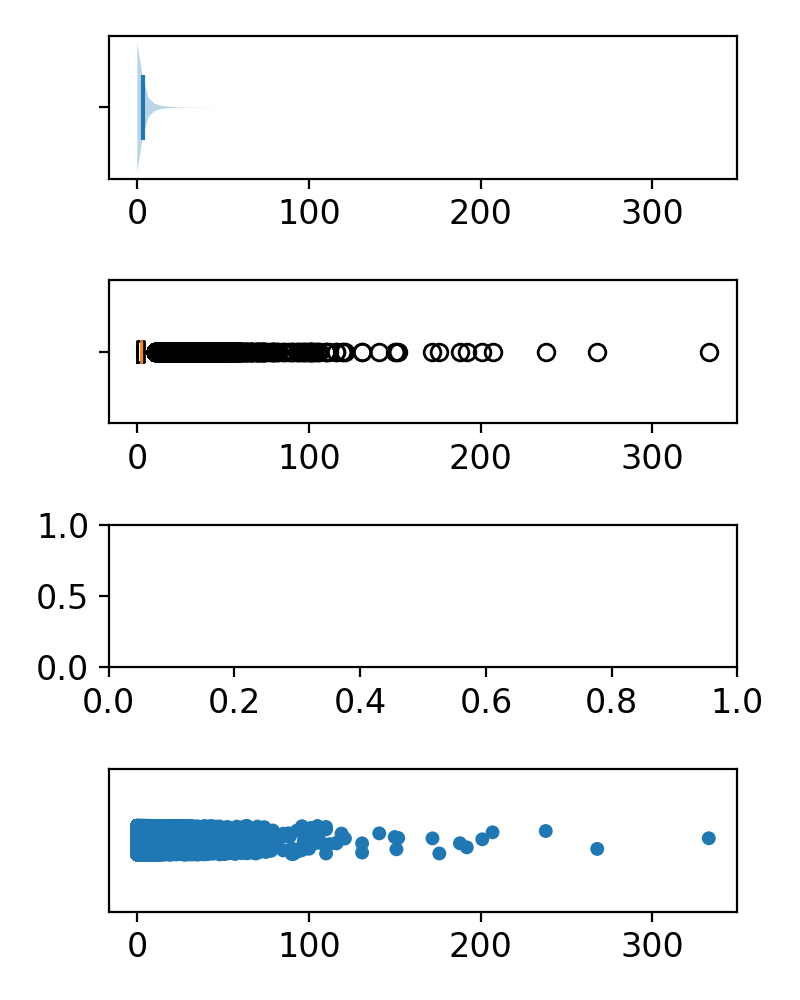}
\plotreal{numberOfCommits}{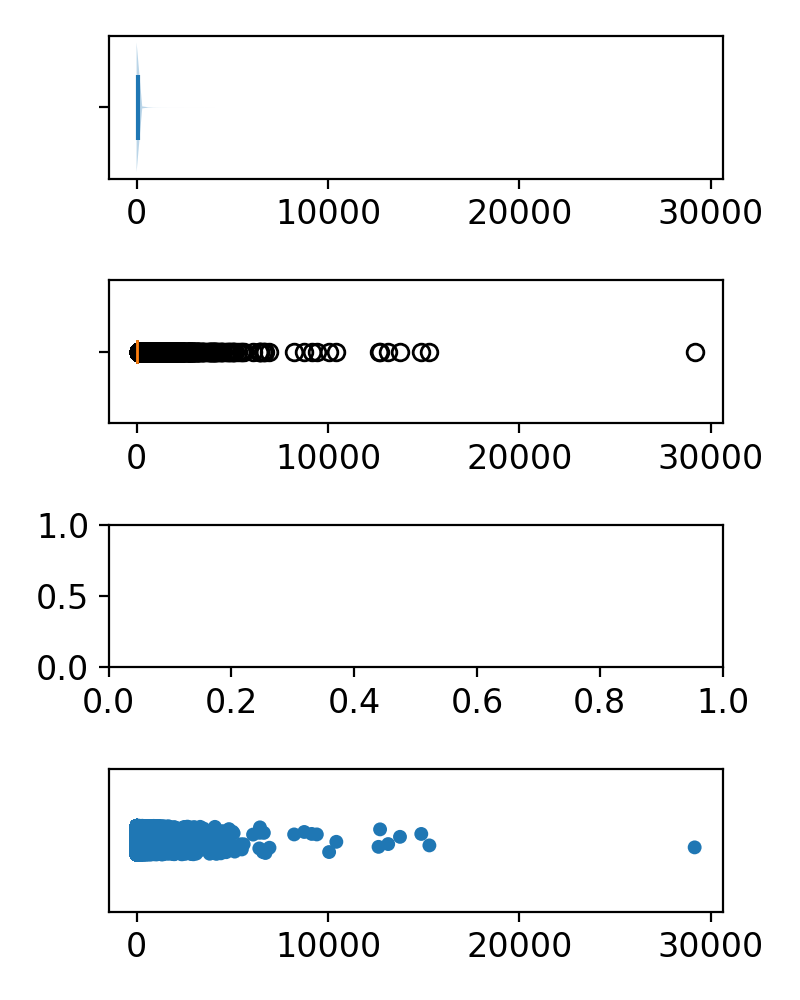}
\plotreal{numberOfContributors}{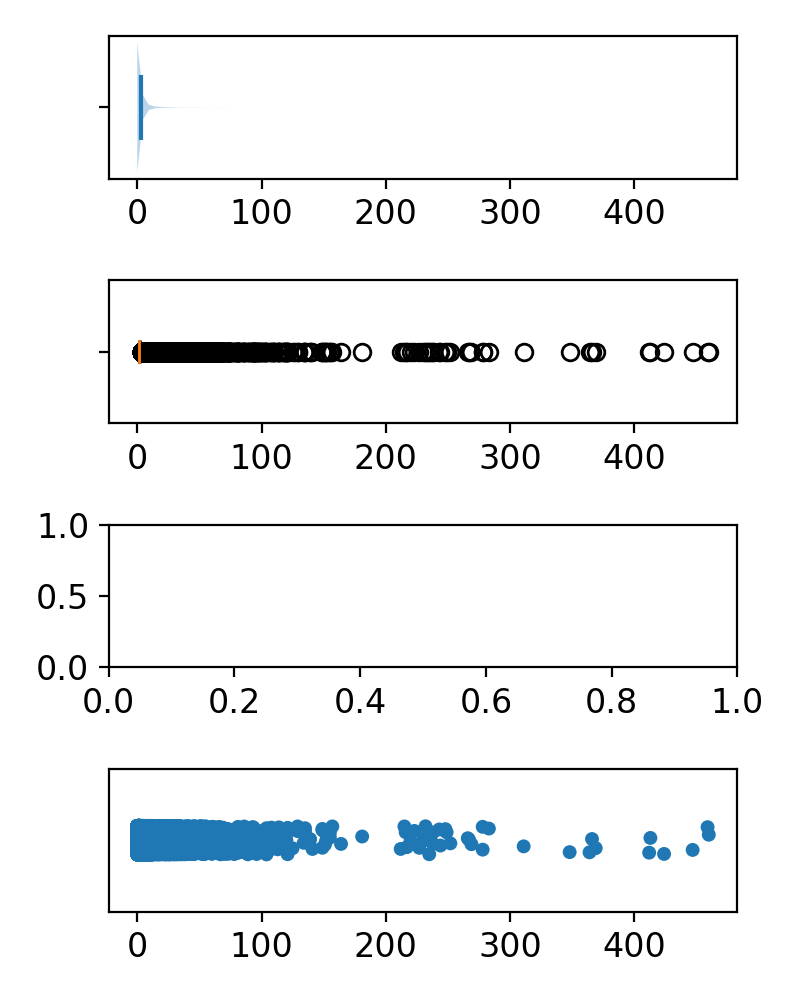}\\
\hspace{8mm}\plotreal{numberOfFiles}{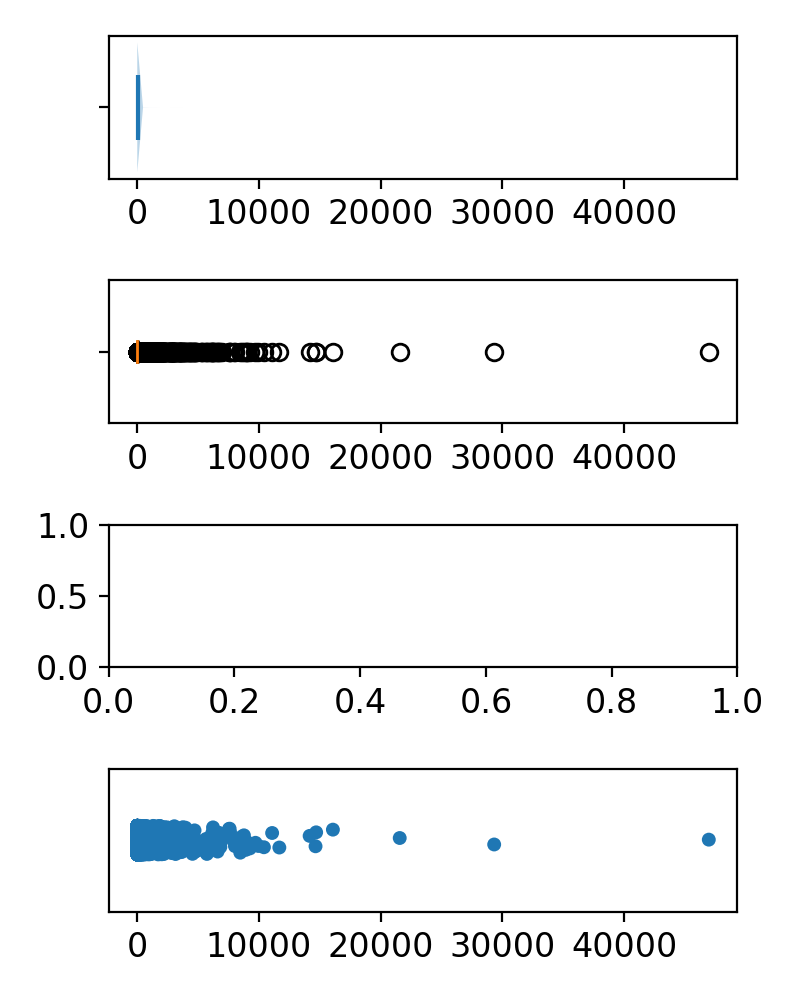}
\plotreal{numberOfHtmlFiles}{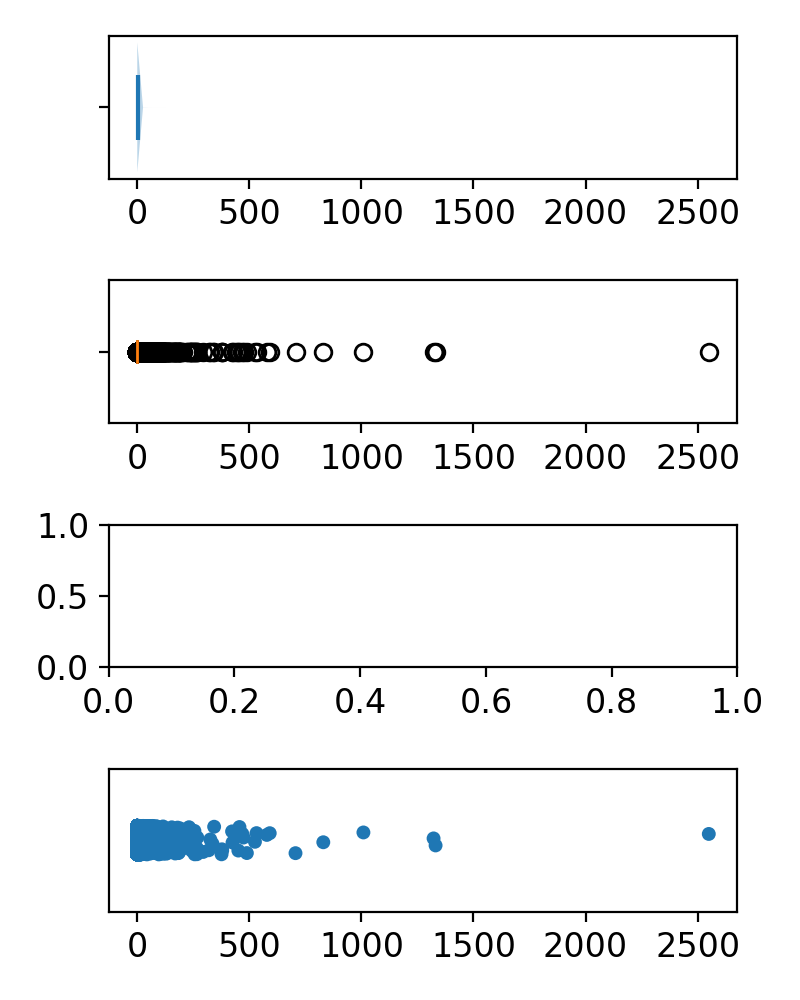}
\plotreal{numberOfIssues}{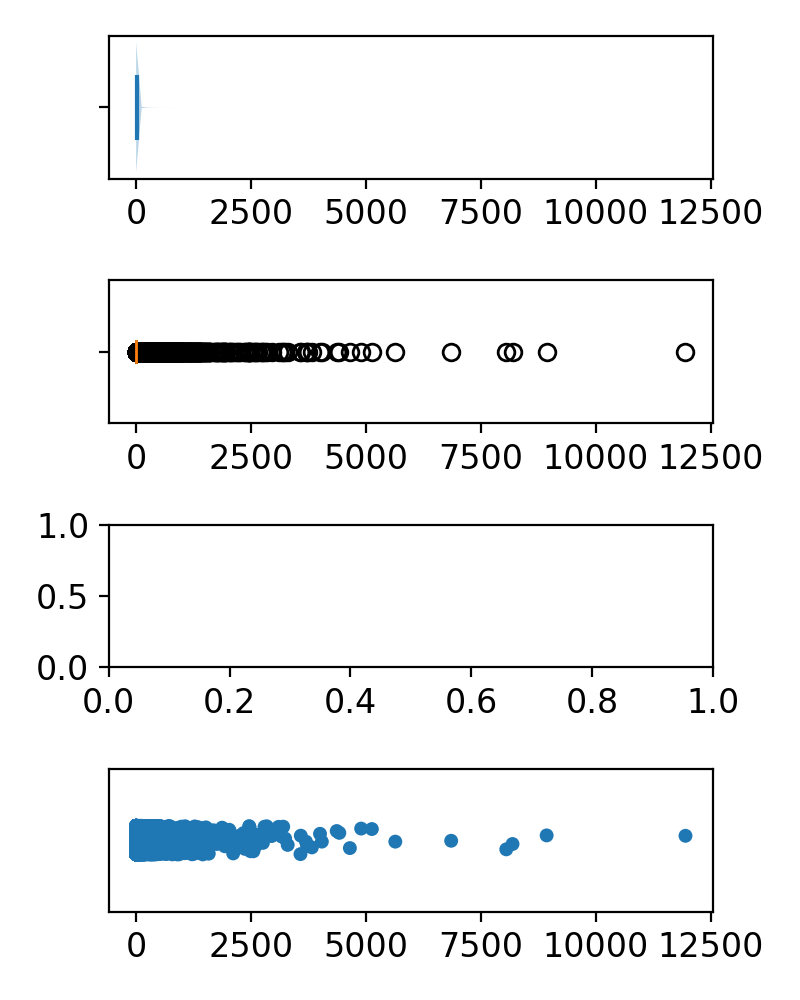}
\plotreal{numberOfJsFiles}{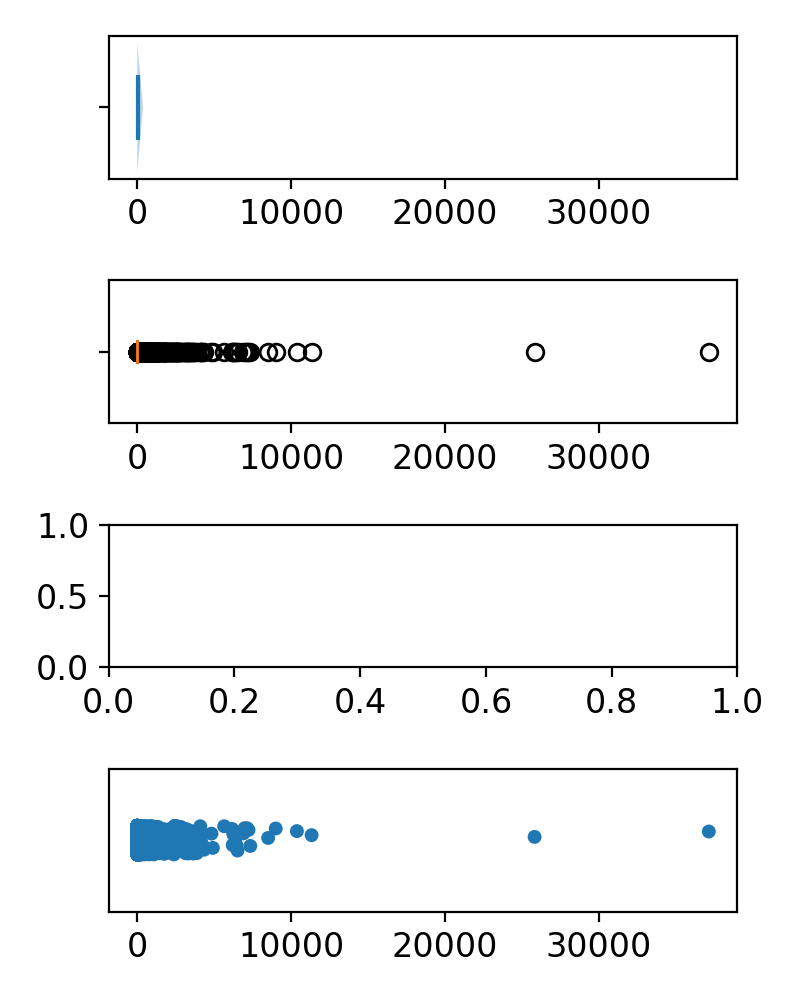}\\\hspace{8mm}\plotreal{numberOfNewcomerLabels}{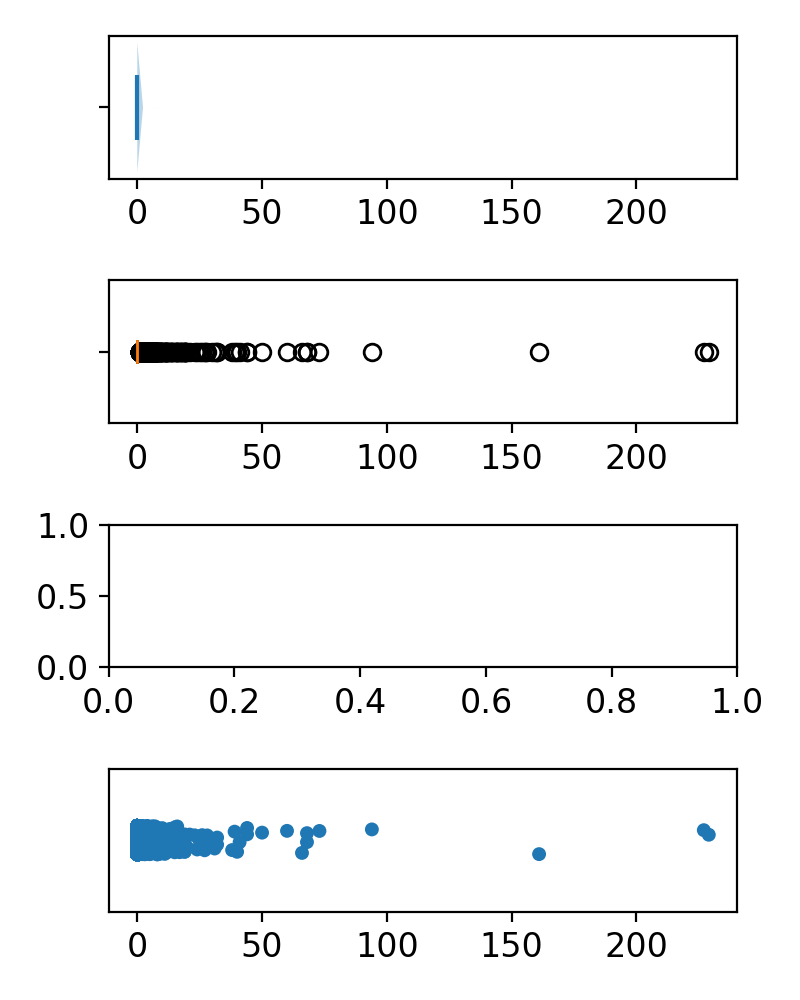}
\plotreal{numberOfPullRequests}{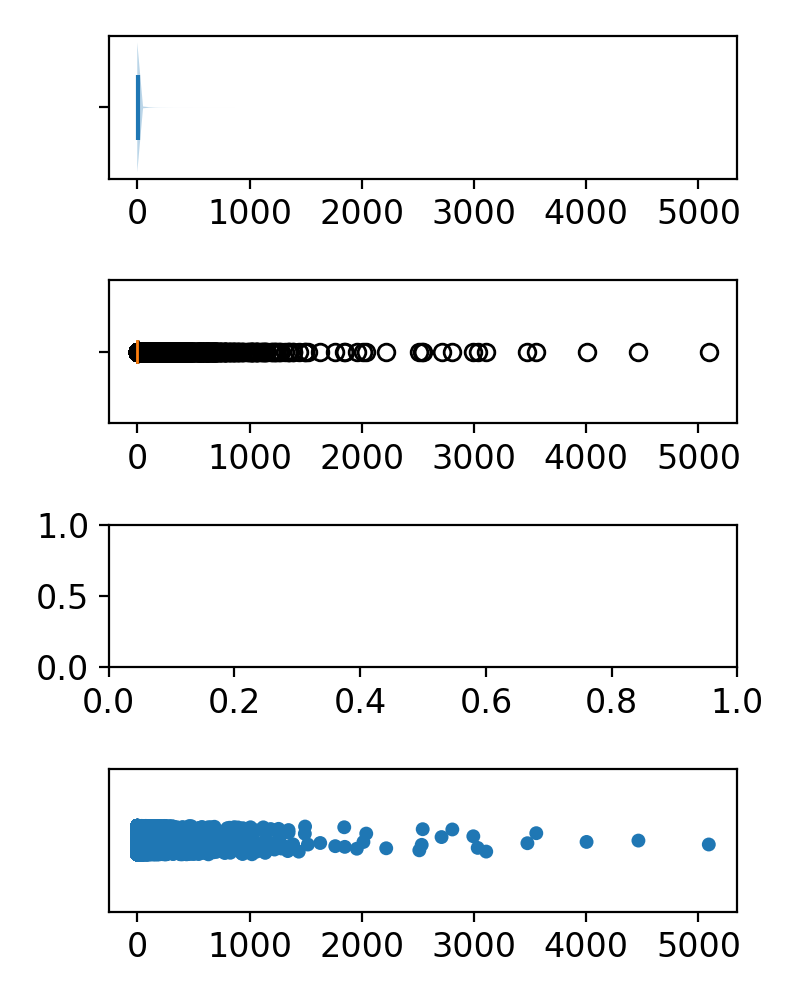}
\plotreal{numberOfRepositoryTags}{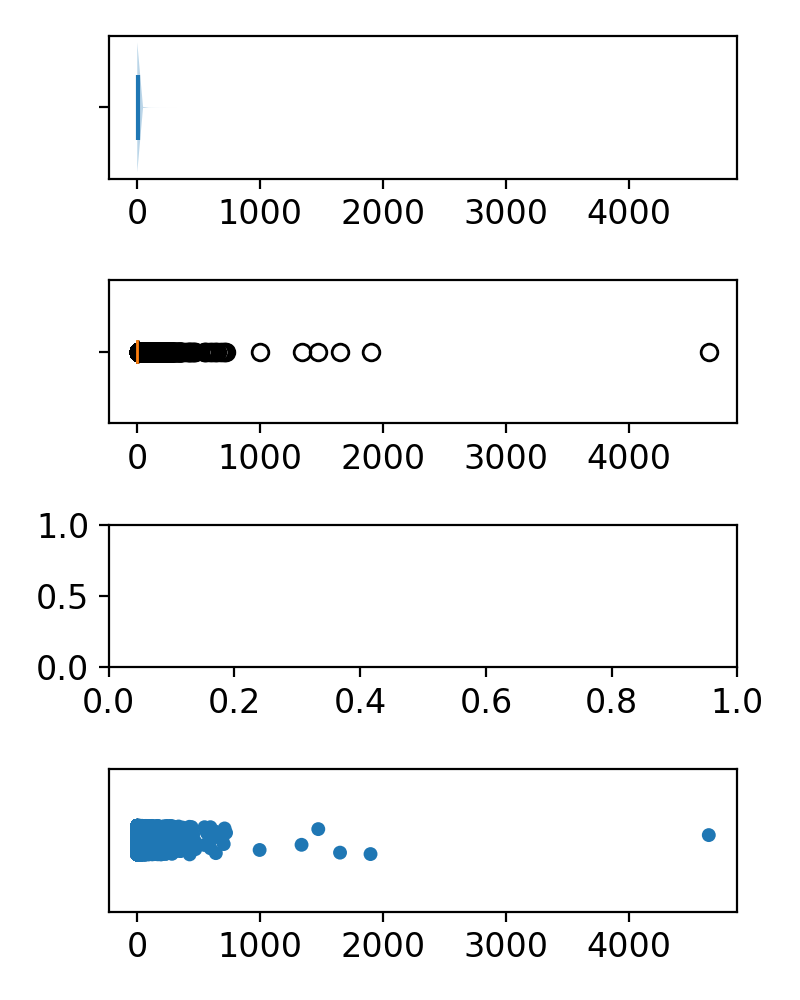}%
\plotreal{readmemdCodeSnippetsNumber}{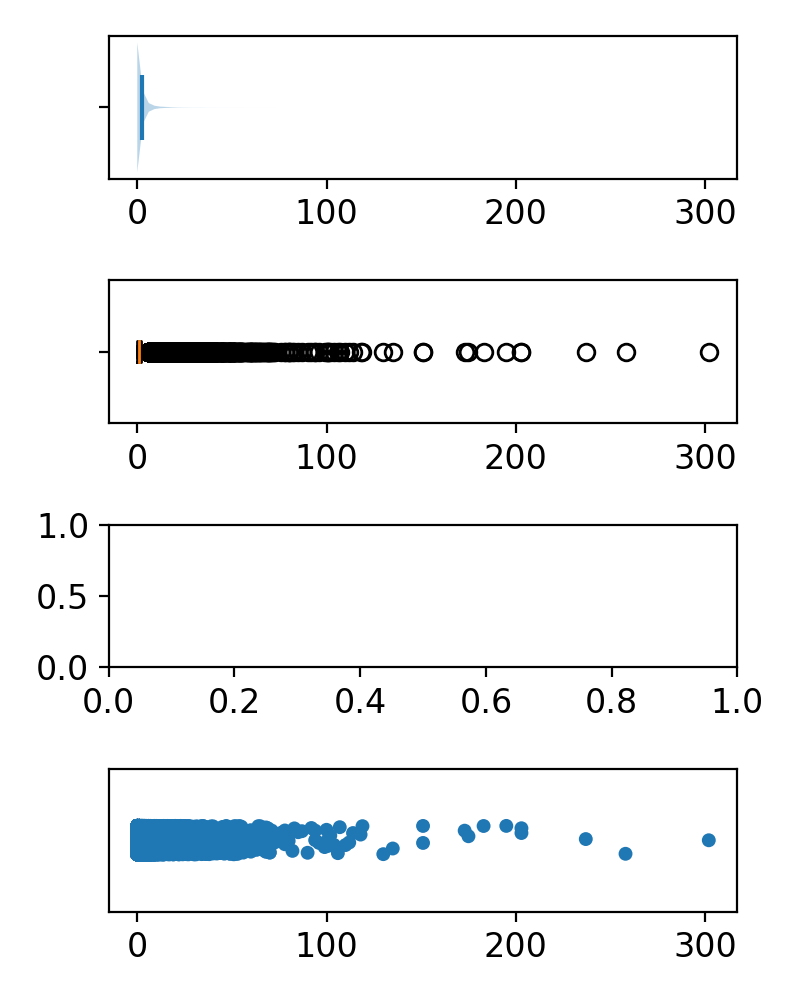}%
\\\hspace{8mm}%
\plotreal{sizeOfRepository}{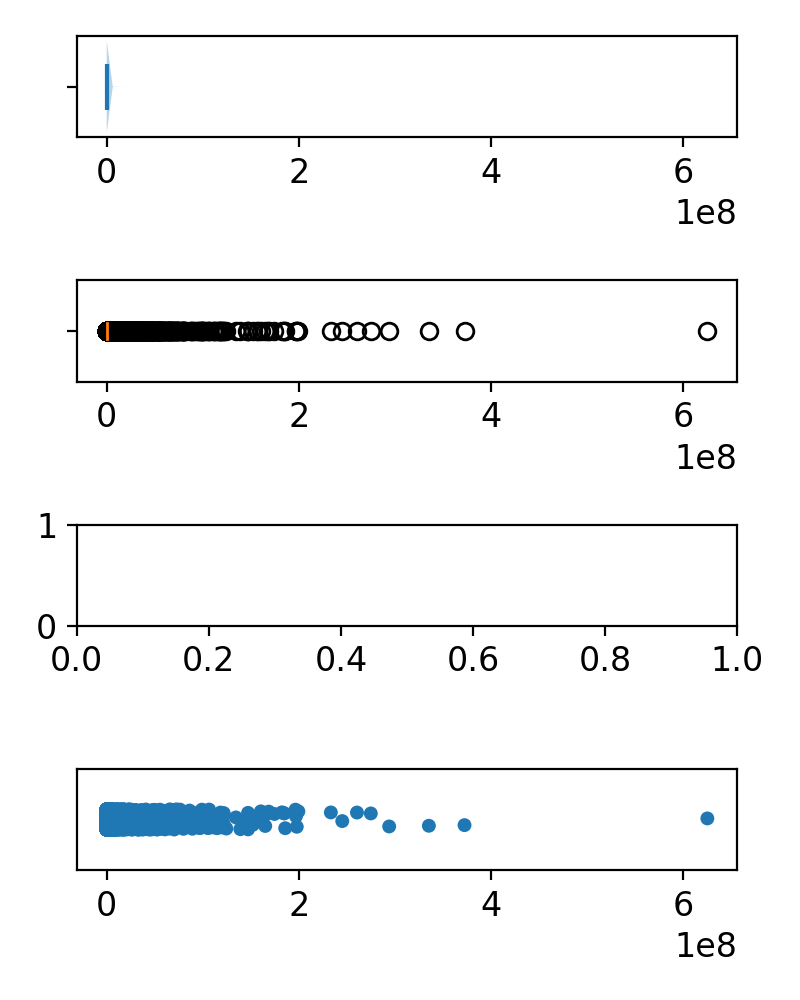}
\vspace{3mm}\\%
\rotatebox{90}{\hspace{-20mm}\textsf{\scriptsize binary features\vspace{2mm}}} \rotatebox{90}{\hspace{-32mm}\rule{41mm}{0.4pt}}\hspace{3mm}
\plotbinary{ableToBuild}{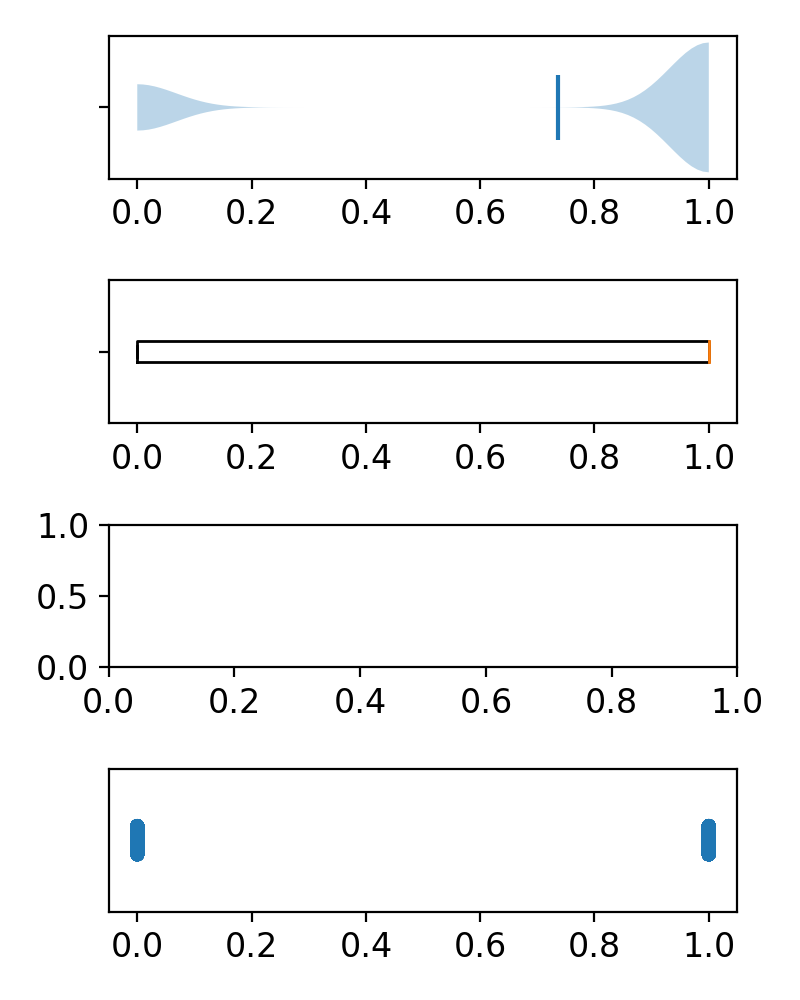}
\plotbinary{ableToExecuteATest}{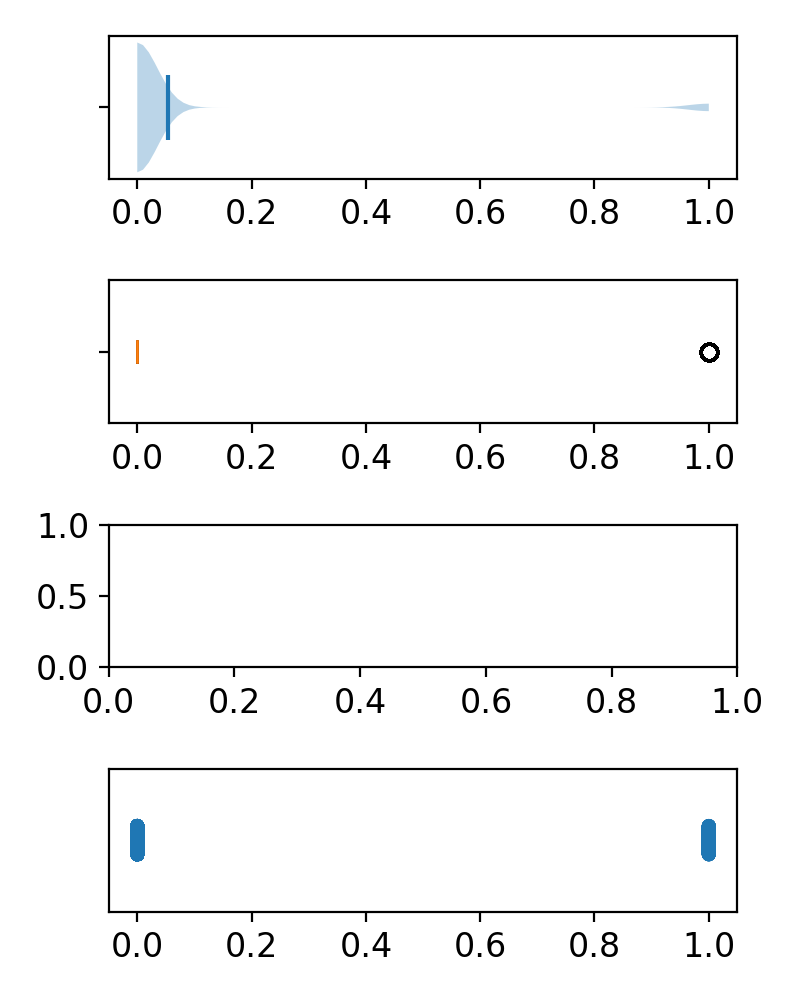}
\plotbinary{ableToInstall}{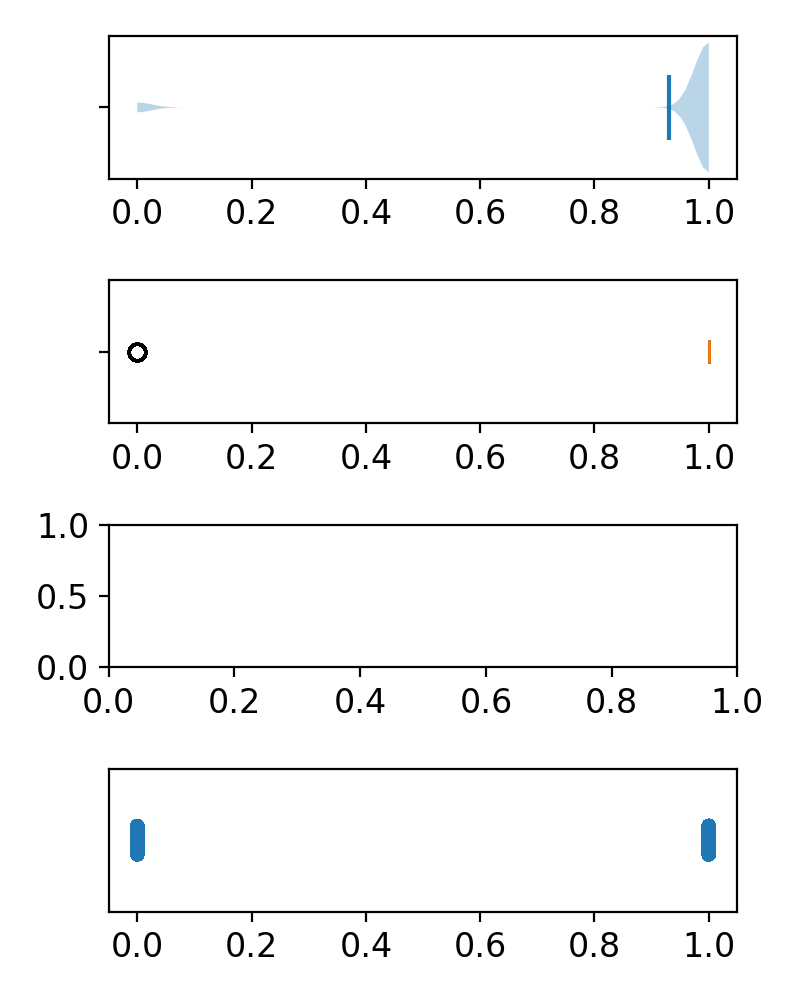}
\plotbinary{hasACodeOfConduct}{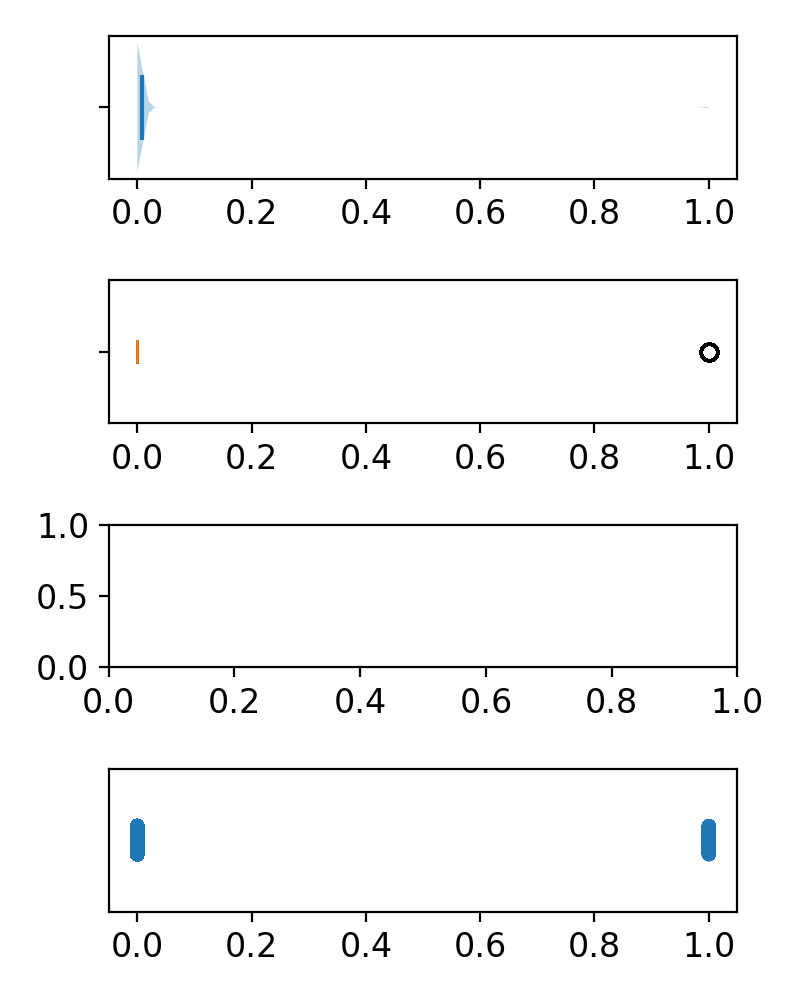}\\
\hspace{8mm}\plotbinary{hasAContributionGuideline}{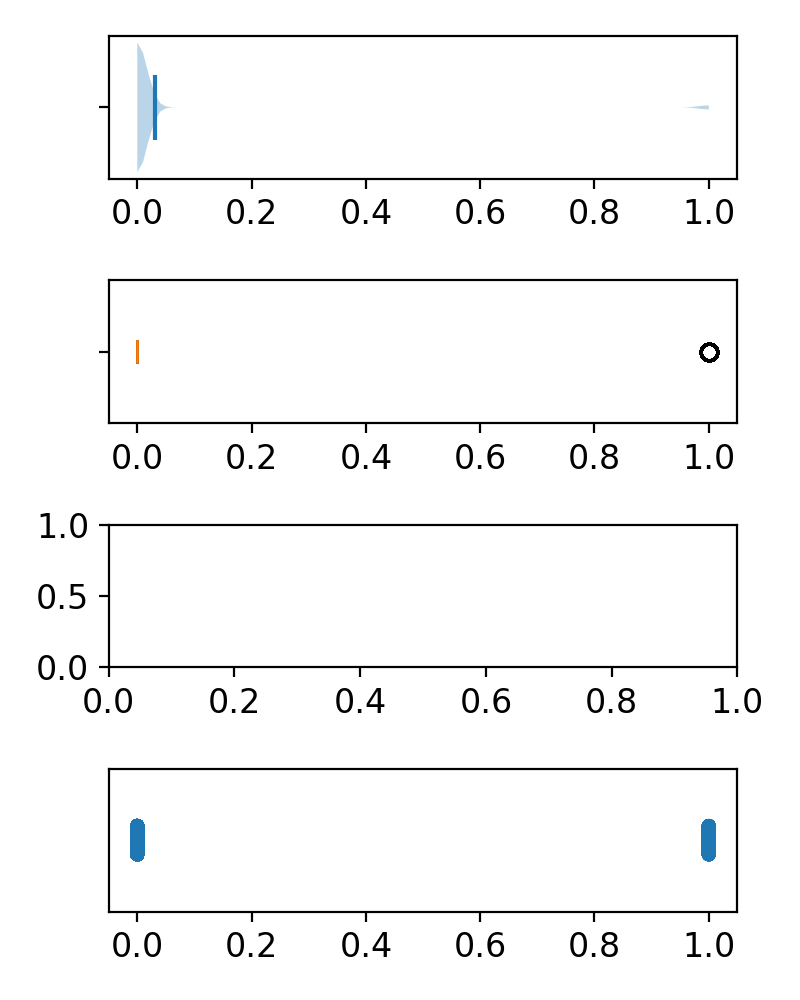}
\plotbinary{hasALicense}{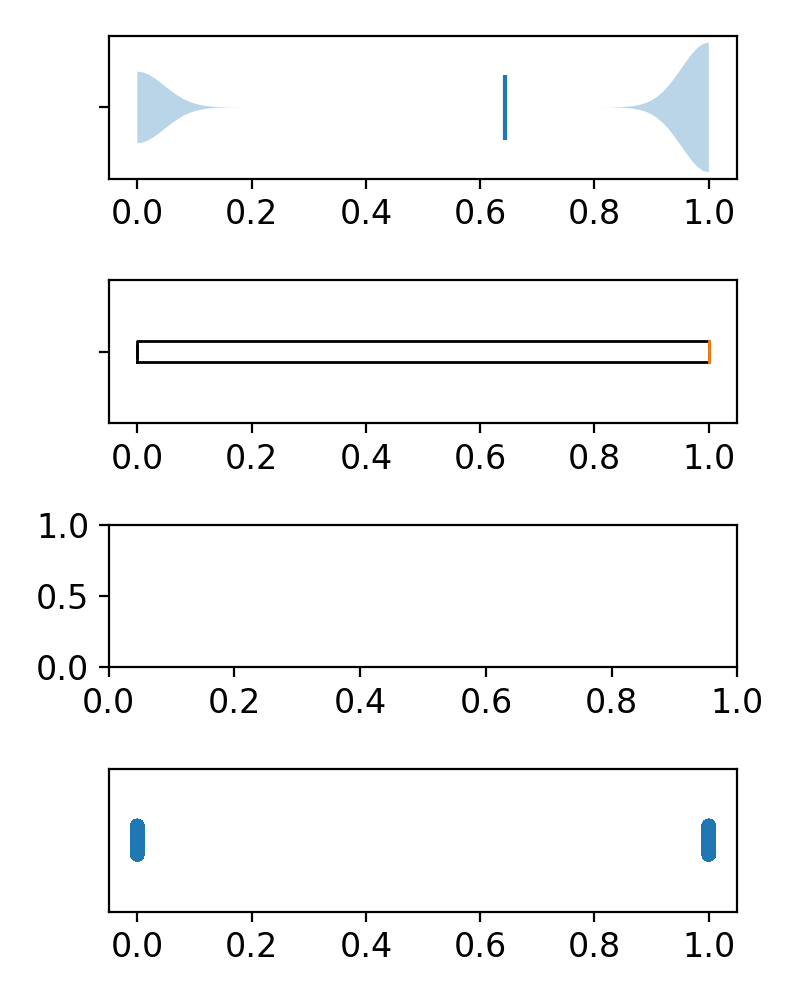}
\plotbinary{hasAnInstallExample}{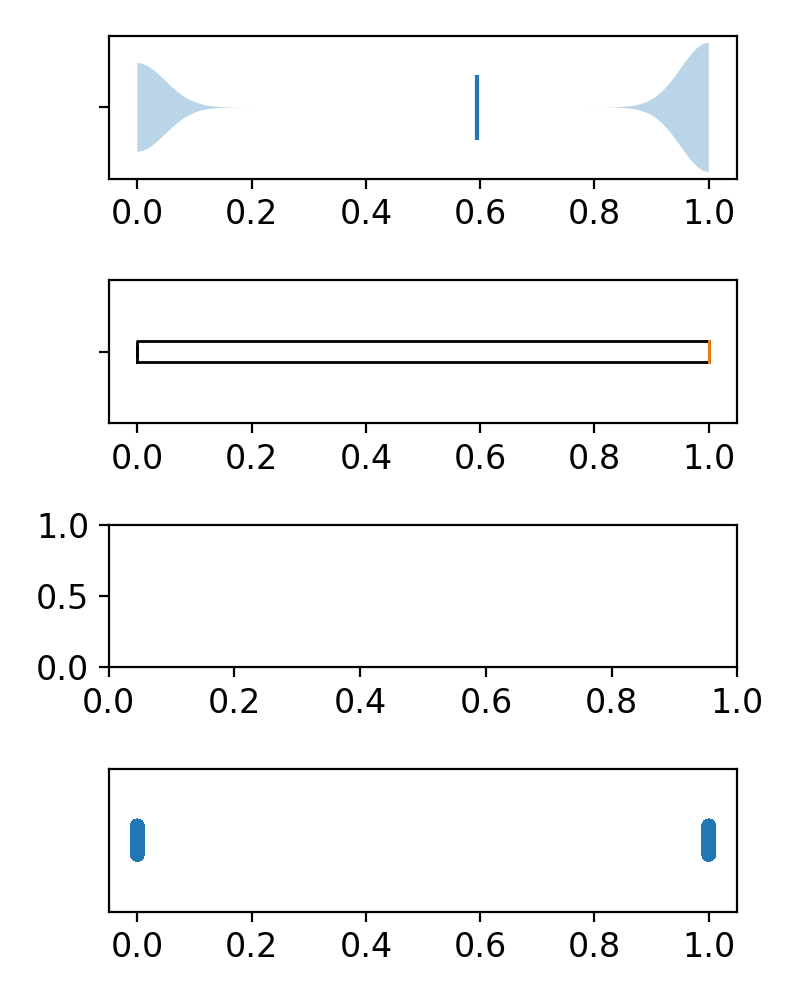}%
\plotbinary{hasAReadmemd}{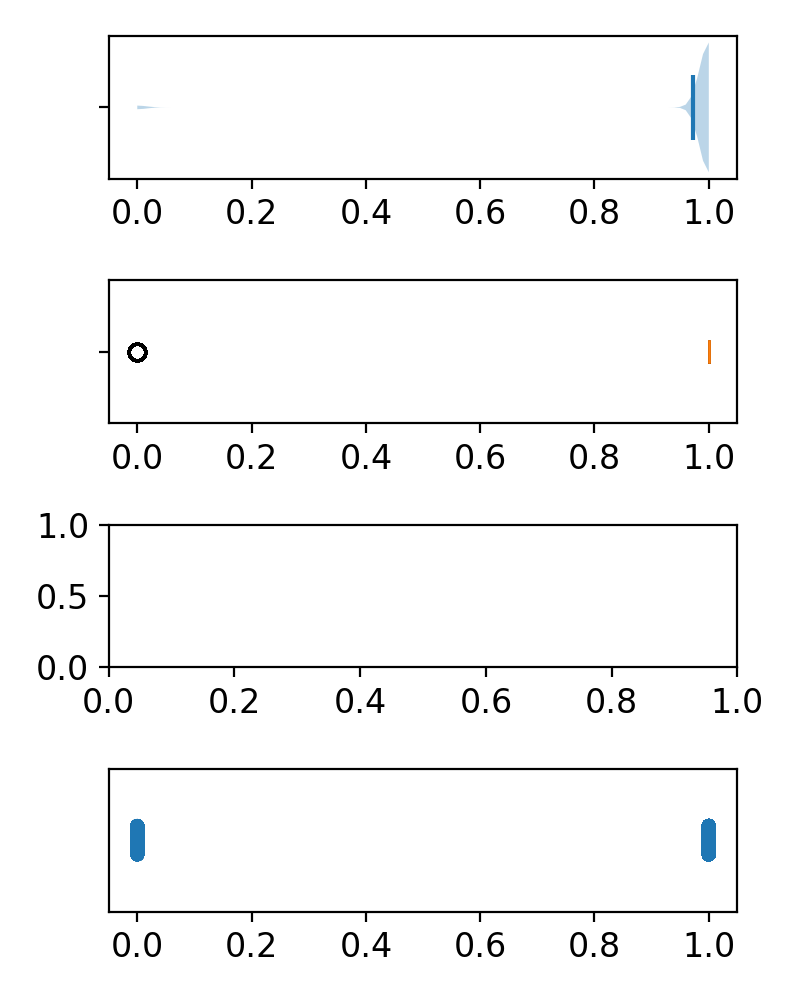}%
\\\hspace{8mm}\plotbinary{hasARunExample}{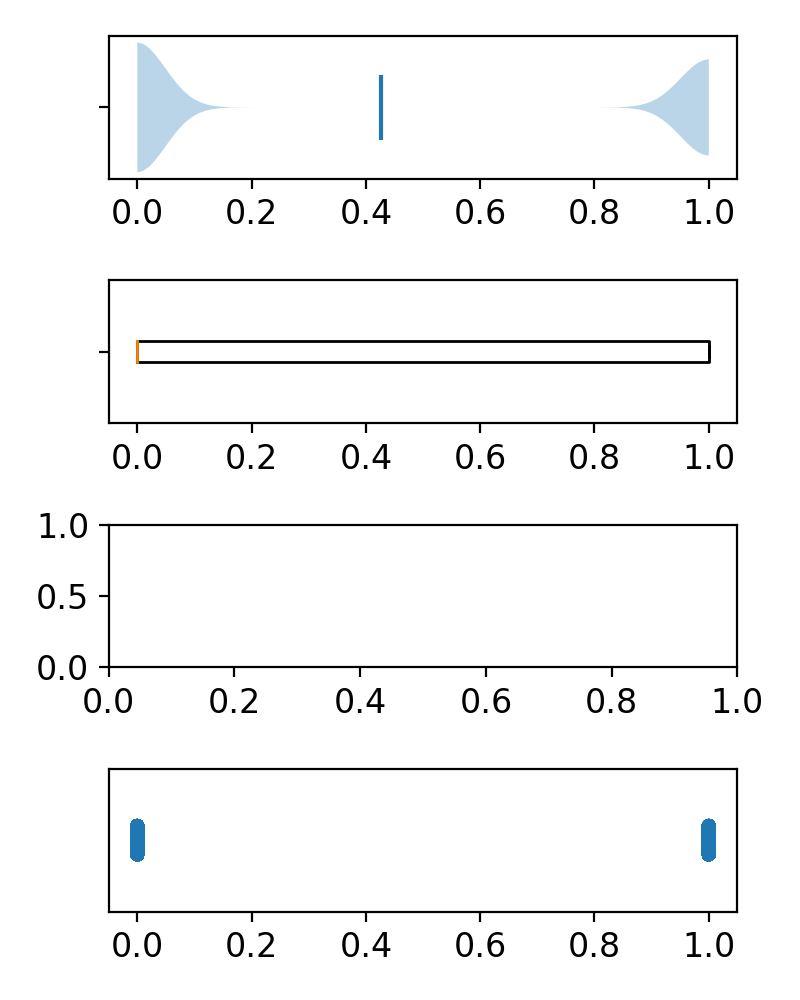}%
\plotbinary{hasASourceDirectory}{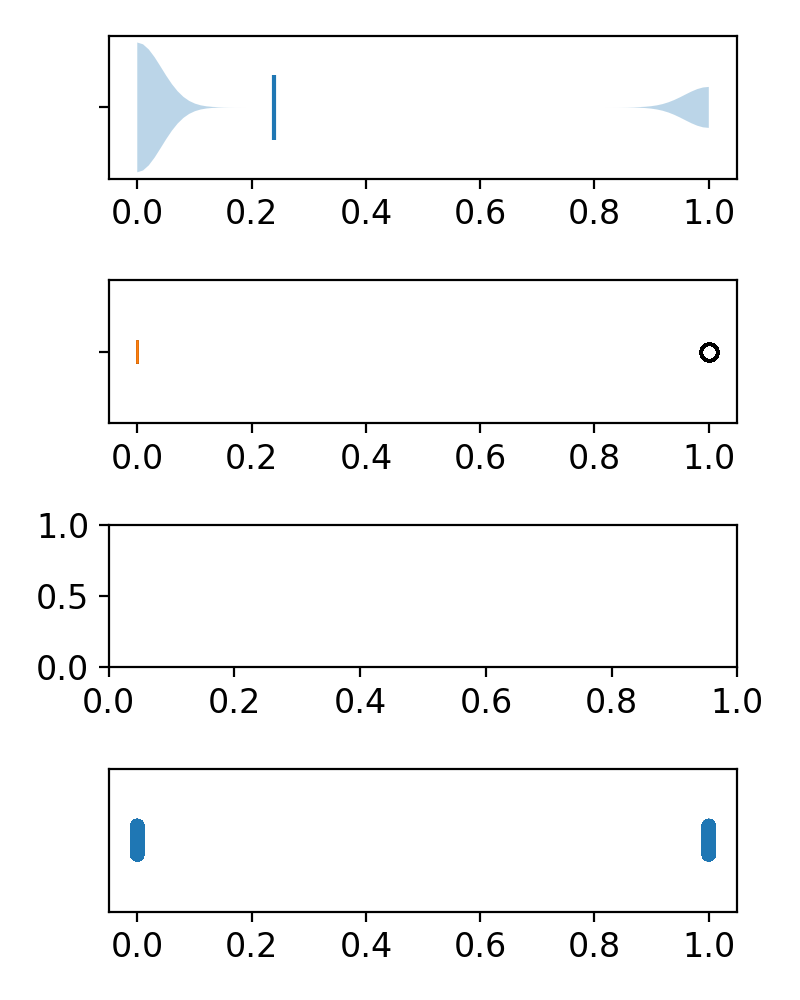}
\plotbinary{hasATestDirectory}{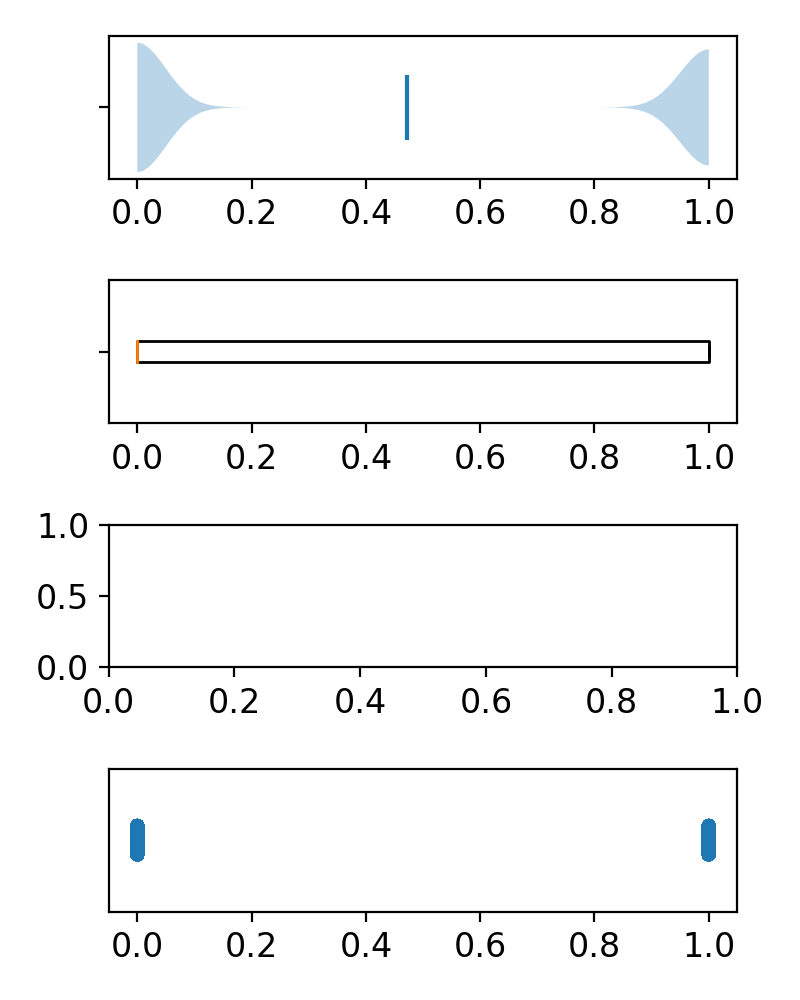}
\caption{Value distribution of 28 features in alphabetical order. For non-binary features (top five rows), we use strip plots to show the distributions qualitatively by showing single data points; for binary features (bottom three rows), we use violin plots with shown means to illustrate the amount of data at either end of the distribution. Shown are all data points (11,127 to 104,364). %
}
\label{fig:repos:plots}%
\end{figure*}

\newcommand{\putlabel}[5]{\node[fill=yellow,fill opacity=0.8,anchor=south,text width=#4mm, inner xsep=1pt,inner ysep=3pt,outer sep=0pt,align=center,text opacity=1, rotate=#3,] at  (#1,#2){\footnotesize #5};}

\begin{figure}[h!]
\begin{minipage}[b]{\linewidth}%
\centering
\begin{tikzpicture}
\node[anchor=south west,inner sep=0] (image) at (0,0) 
{\includegraphics[width=80mm,trim={60 50 50 50},clip]{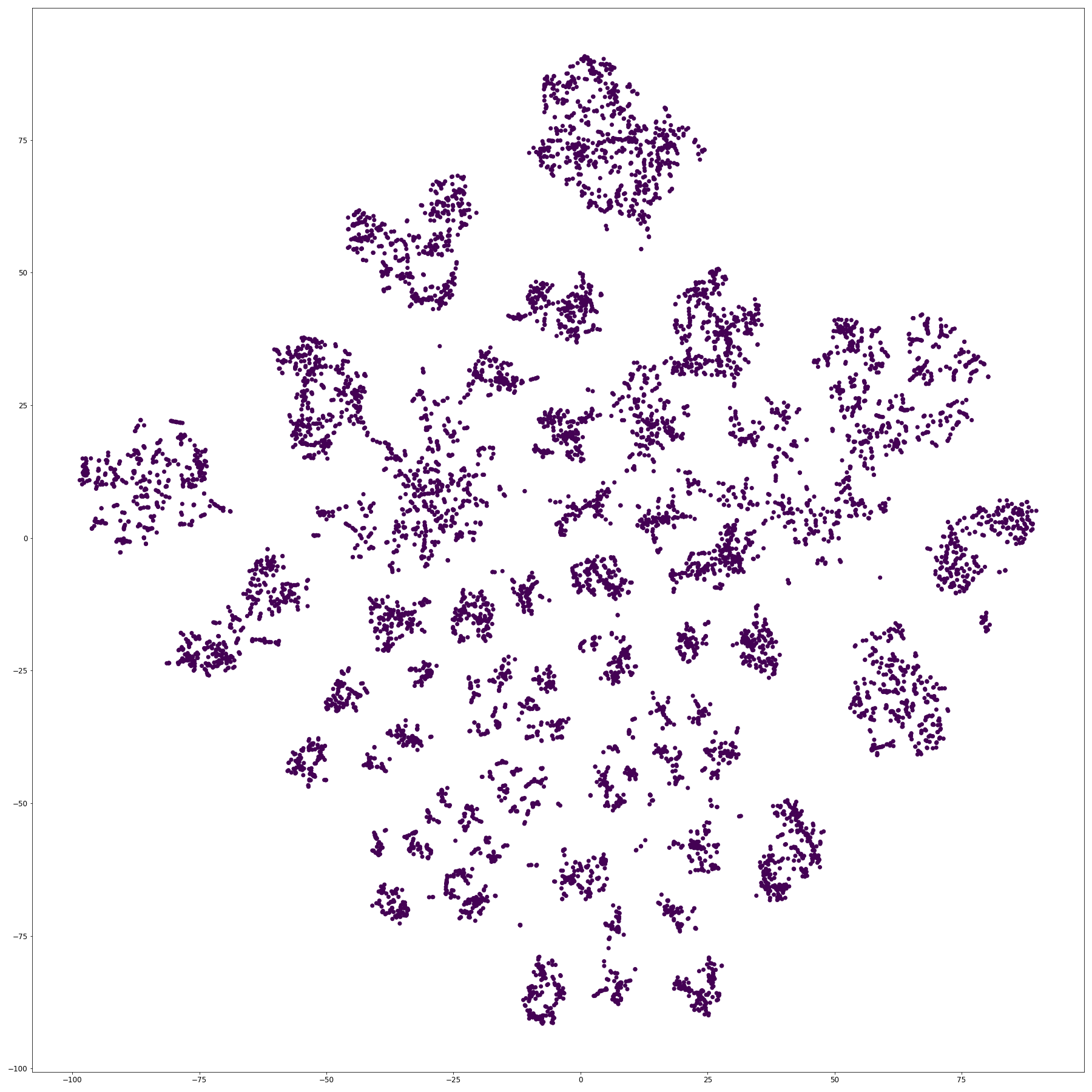}};

\putlabel{6.6}{2.6}{-60}{18}{!ableToInstall}
\putlabel{5.6}{3.1}{60}{18}{!ableToBuild}
\putlabel{.8}{4.6}{60}{25}{ableToExecuteATest}
\putlabel{6.3}{5.3}{-65}{34}{hasAContributionGuideline}
\putlabel{7.5}{4.2}{-90}{28}{hasACodeOfConduct}
\putlabel{3.2}{2.9}{15}{28}{very low readmemdCodeSnippetsNumber}
\putlabel{3.2}{1.5}{30}{20}{!hasALicense}
\putlabel{3.7}{5.2}{9}{28}{hasAnInstallExample}
\putlabel{4.0}{6.5}{20}{30}{large numberOfIssues}
\end{tikzpicture}
\end{minipage}%
\caption{Repositories in 2D. The axes do not have any particular meaning in projections like these, which is why we removed them.
Note that the clusters at the very bottom end of this figure are not easily characterised by a single feature but by combinations of features.}%
\label{fig:repos:tsne}\vspace{-2mm}%
\end{figure}

\section{Correlating features by Perspective}

\begin{table}[]
\centering
\caption{Top 5 negative correlations of features grouped by their types (U: user, C: contributor, B: both of them, N: none of them).}
\label{tab:correlation}
\scalebox{0.9}{
\begin{tabular}{@{}cccl@{}}
\toprule 
\textbf{Feature} & \textbf{Software} & \multirow{2}{*}{\textbf{Correlation}} & \multicolumn{1}{c}{\textbf{Corresponding}} \\
\textbf{type} & \textbf{feature} & & \multicolumn{1}{c}{\textbf{feature}} \\ 
\midrule
\multirow{14}{*}{\rotatebox{90}{Documentation}} & \multirow{5}{*}{ableToBuild}& {-0.0962} & {hasAContributionGuideline} (C) \\
 & & -0.0536 & hasACodeOfConduct (C) \\
 & & -0.0448 & linesInReadmemd (B) \\
 & & -0.0426 & numberOfCodeBlocks (U) \\
 & & -0.0393 & hasARunExample (U) \\ \cmidrule(l){2-4}
 & \multirow{5}{*}{ableToExecuteATest} & {-0.0518} & {hasAContributionGuideline} (B) \\
 & & -0.0281 & hasACodeOfConduct (C) \\
 & & -0.0199 & hasARunExample (U) \\
 & & -0.0074 & hasAReadmemd (B) \\
 & & -0.0065 & numberOfCodeBlocks (U) \\ \cmidrule(l){2-4}
 & \multirow{4}{*}{ableToInstall} & {-0.0255} & {hasACodeOfConduct} (C) \\
 & & -0.0216 & hasAContributionGuideline (C) \\
 & & -0.0096 & hasAnInstallExample (B) \\
 & & -0.0092 & linesInReadmemd (B) \\
\midrule 
\multirow{15}{*}{\rotatebox{90}{Repository}} & \multirow{5}{*}{ableToBuild} & {-0.2485} & {numberOfFiles} (N) \\
 & & -0.2291 & numberOfCommits (N) \\
 & & -0.2224 & sizeOfRepository (B) \\
 & & -0.2126 & hasASourceDirectory (C) \\
 & & -0.1847 & numberOfRepositoryTags (N) \\ \cmidrule(l){2-4}
 & \multirow{5}{*}{ableToExecuteATest} & {-0.1141} & {sizeOfRepository} (B) \\
 & & -0.1114 & numberOfCommits (N) \\
 & & -0.0997 & numberOfFiles (N) \\
 & & -0.0897 & hasASourceDirectory (C) \\
 & & -0.0818 & numberOfRepositoryTags (N) \\ \cmidrule(l){2-4}
 & \multirow{5}{*}{ableToInstall} & {-0.0947} & {sizeOfRepository} (B) \\
 & & -0.0808 & numberOfFiles (N) \\
 & & -0.0683 & numberOfCommits (N) \\
 & & -0.0648 & numberOfJsFiles (N) \\
 & & -0.0270 & numberOfRepositoryTags (N) \\ 
\midrule
\multirow{12}{*}{\rotatebox{90}{GitHub activity}} & \multirow{4}{*}{ableToBuild} & {-0.2249} & {numberOfIssues} (B) \\
 & & -0.2139 & numberOfPullRequests (B) \\
 & & -0.1893 & numberOfContributors (B) \\
 & & -0.0662 & numberOfNewcomerLabels (C) \\ \cmidrule(l){2-4}
 & \multirow{4}{*}{ableToExecuteATest} & {-0.1099} & {numberOfIssues} (B) \\
 & & -0.1064 & numberOfContributors (B) \\
 & & -0.1029 & numberOfPullRequests (B) \\
 & & -0.0235 & numberOfNewcomerLabels (C) \\ \cmidrule(l){2-4}
 & \multirow{4}{*}{ableToInstall} & {-0.0484} & {numberOfIssues} (B) \\
 & & -0.0466 & numberOfContributors (B) \\
 & & -0.0379 & numberOfPullRequests (B) \\
 & & -0.0274 & numberOfNewcomerLabels (C) \\
\bottomrule
\end{tabular}
}
\end{table}

To answer \RqOneText~we initially perform a descriptive analysis of all features. 
From this analysis, we first describe the most positive and most negative correlations in the dataset to understand the trade-off between features and perspectives.
We then manually investigate correlations between different perspectives.
The following analysis characterises the 11,127 packages for which all 30 feature values are available; we ignore githubLink and mostPopularFileExtension. %
This was necessary to achieve a consistent picture, as not all used techniques allow for missing values.

Figure~\ref{fig:repos:instanceclusters} shows the Spearman rank-order correlation coefficients between the 28 features and clustered with Wards hierarchical clustering approach.
We find that features from the same type are more likely to have strong positive correlations among them.
The GitHub activity features like numberOfIssues and numberOfPullRequests have the strongest positive correlation among any feature pairs.
For documentation features, numberOfCodeBlocks and linesInReadmemd have a strong positive correlation, but less than the GitHub activity features.
For repository features, numberOfCommits and numberOfRepositoryTags also have a strong positive correlation and belong to the same cluster.
Software features (i.e., ableToExecuteATest, ableToInstall, ableToBuild) have positive correlations among each other, but the correlations are not strong.
On the other hand, the features from different types tend to have either small positive correlations or negative correlations.

Table~\ref{tab:correlation} shows the top negative correlations between software features and other features, which are top negative correlations among any feature pairs shown in Figure~\ref{fig:repos:instanceclusters}.
We find that various features that measure the size of a repository (e.g., number of files, commits, issues, pull requests, contributors) tend to be negatively correlated with the ability to install, build, and execute tests on a package.
This makes intuitive sense as larger repositories are more complex and more likely to encounter issues regarding runnability. 
In addition, we note that correlations between several of the documentation features and runnability also tend to be slightly negative.
Curiously, even the correlation between hasAnInstallExample and the ability to install a package is not positive.

Figure~\ref{fig:repos:plots} shows the distributions of the feature values. 
Note that the strip plots cannot show all 11,127 data points; however, they provide a subjectively better qualitative and quantitative characterisation of the distribution than violin plots.

Figure~\ref{fig:repos:tsne} shows the clusters of npm packages using the t-distributed Stochastic Neighbour Embedding (t-SNE)~\cite{Maaten2008tsne} to project the 28D data points into 2D. 
t-SNE’s reduction process attempts to preserve the distances in the high-dimensional space as much as possible. 
Interestingly, a large number of clusters has formed. 
We have manually inspected the clusters and added labels for cases that we have found interesting. 
While we have to be careful not to over-interpret this reduced representation, we can observe co-located areas where the neighbourhood seems intuitively reasonable. For example, at the right, repositories are listed with contribution guidelines and codes of conduct, both of which can be qualities of mature repositories -- and we can indeed find these repositories in the vicinity of those with large numbers of issues; further to the left, we can find repositories for which it is possible to execute tests. Similarly, we can find repositories that cannot be built or do not have a license in the lower part of the plot. Lastly, and a little bit to the left of the centre of the figure, we have repositories where the README.md has a small number of code snippets: this seems to place them at the (fuzzy) boundary between possible immature projects and mature ones.

\begin{tcolorbox}
    \textbf{Summary:} There are strong positive correlations among features from the same type. Other feature combinations present trade-offs -- in particular software features tend to be negatively correlated with other features. 
\end{tcolorbox}

\begin{figure}[]
\centering%
\begin{subfigure}[b]{0.30\textwidth}
    \centering
    \includegraphics[height=50mm,trim={570 0 0 0},clip]{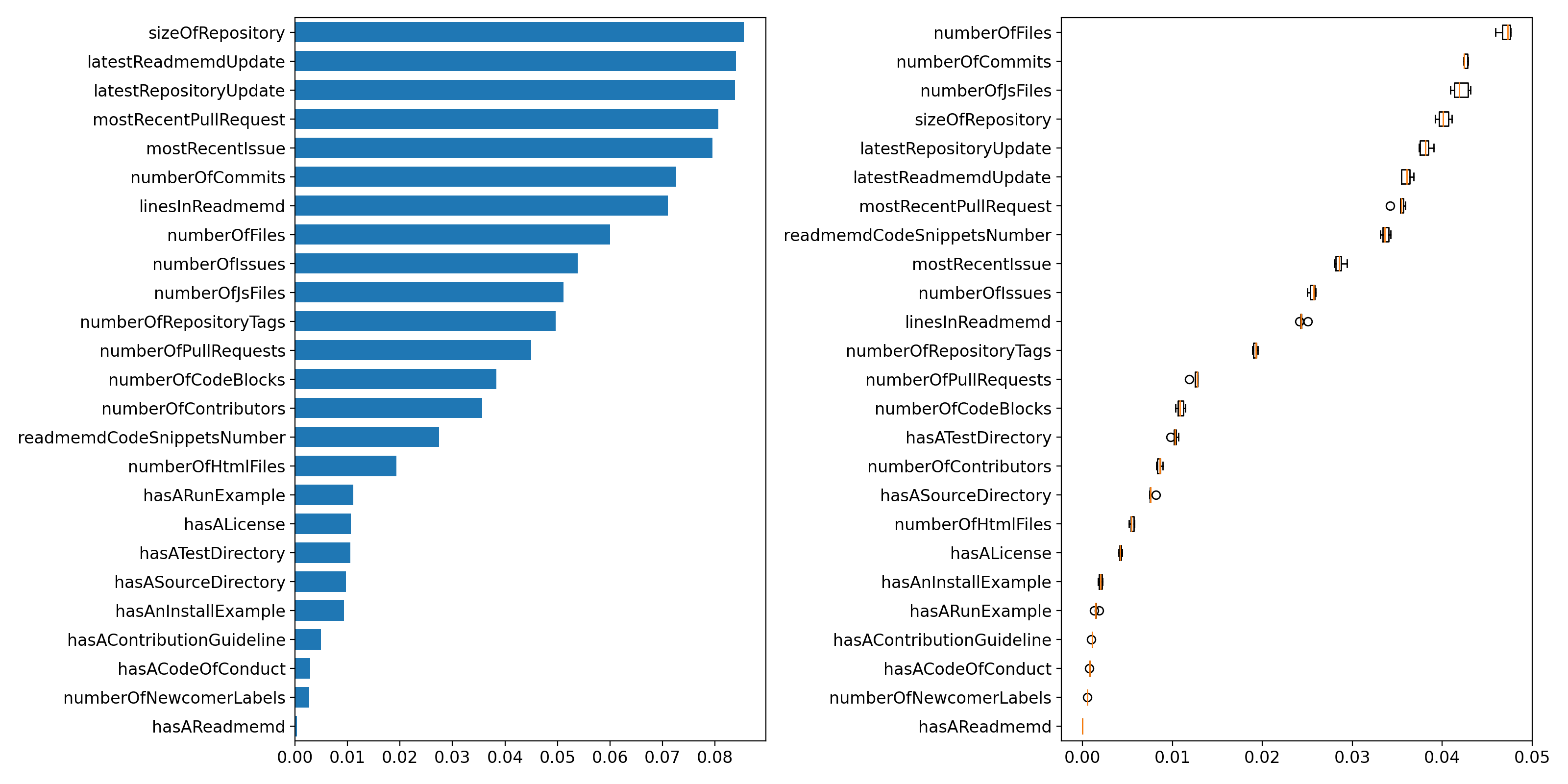}%
    \caption{ableToInstall}
\end{subfigure}
\hfill
\begin{subfigure}[b]{0.30\textwidth}
    \centering
    \includegraphics[height=50mm,trim={582 0 0 0},clip]{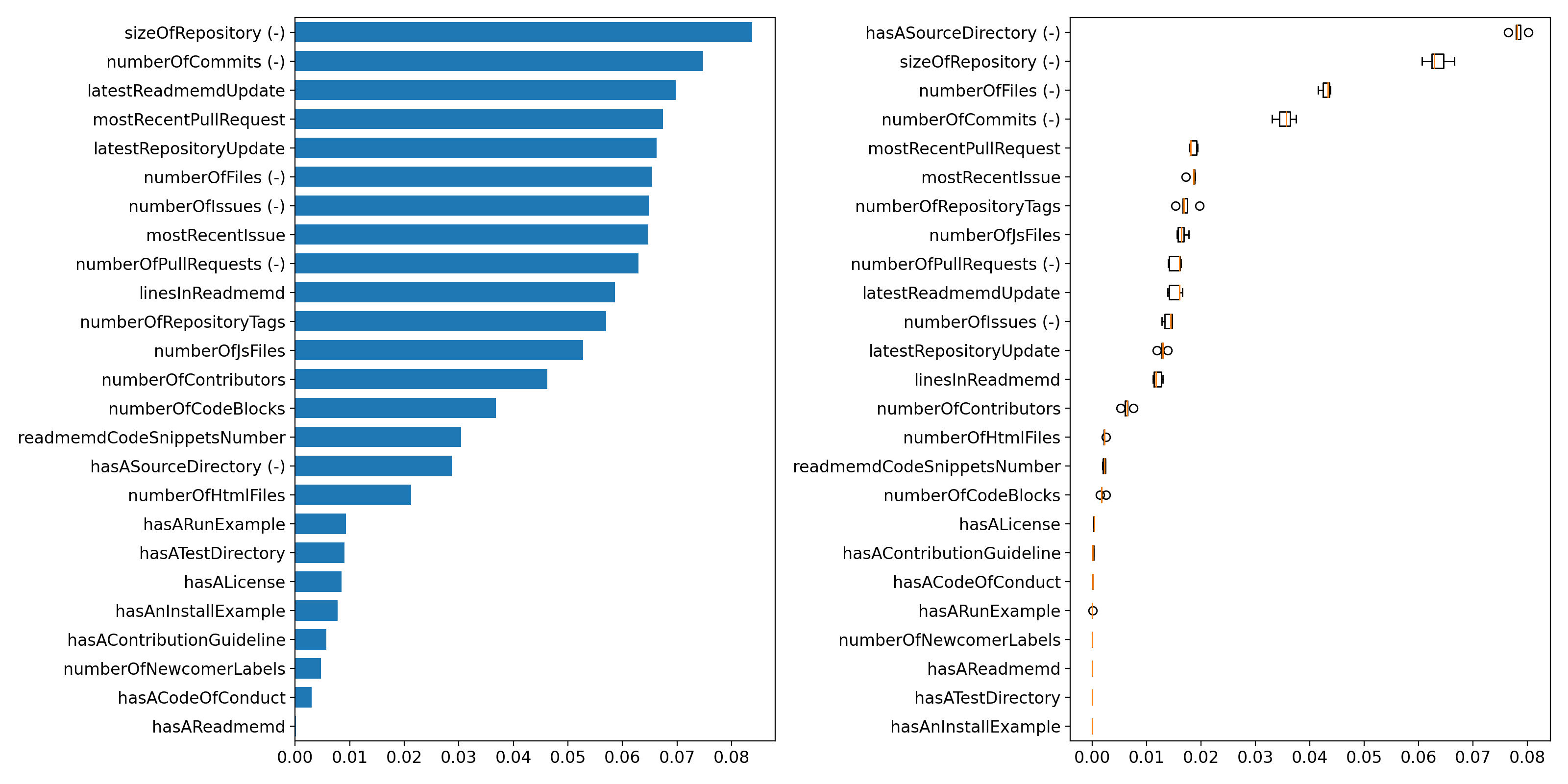}%
    \caption{ableToBuild}
\end{subfigure}
\hfill
\begin{subfigure}[b]{0.30\textwidth}
    \centering
    \includegraphics[height=50mm,trim={582 0 0 0},clip]{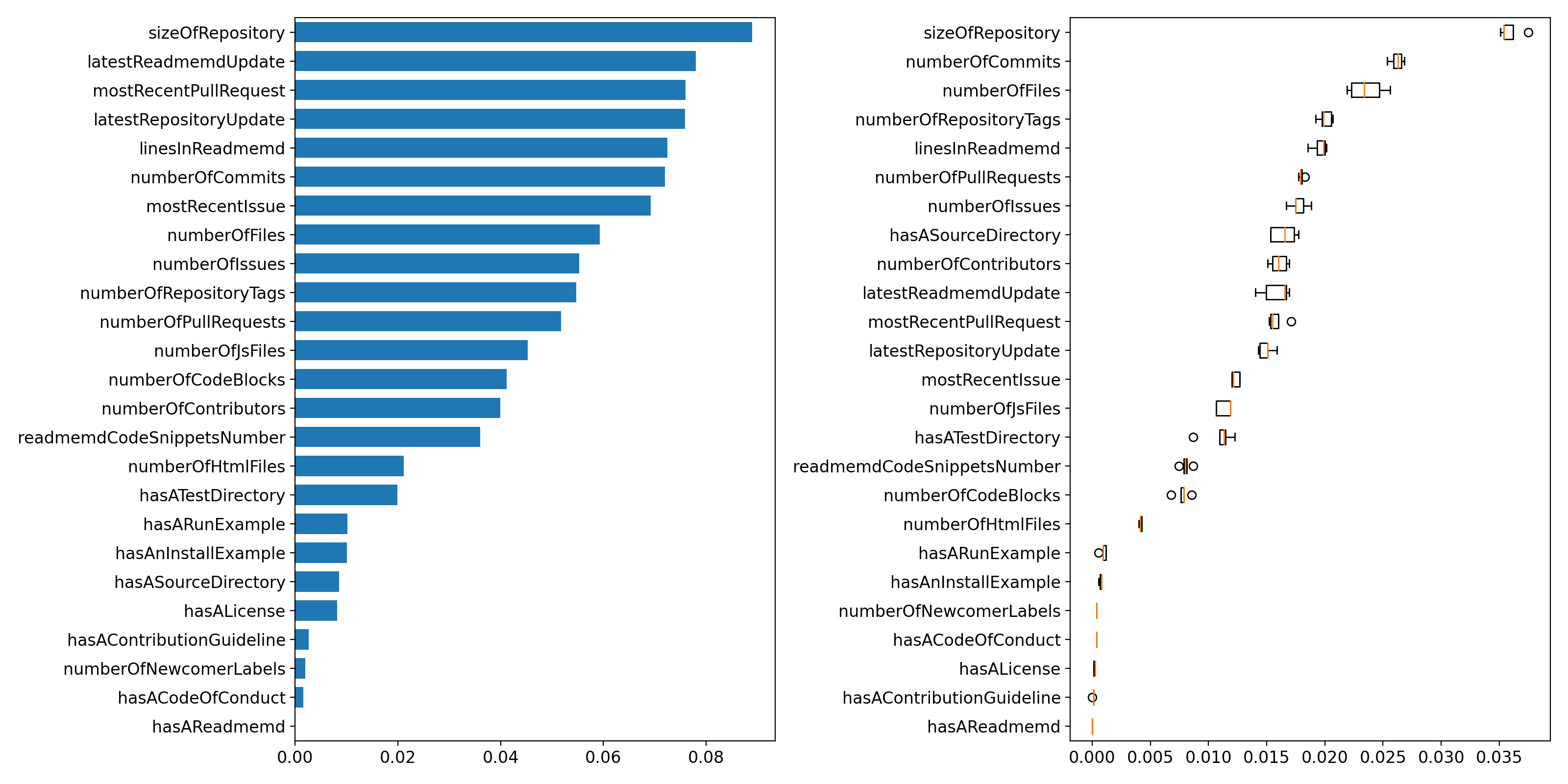}%
    \caption{ableToExecuteATest}
\end{subfigure}
\caption{Permutation importance of a good predictive model for ableToInstall, ableToBuild, and ableToExecuteATest (from left to right). Larger values mean that the prediction is more sensitive with respect to that parameter; hence it can be seen as more important. The $+$ and $-$ mark features with weak correlations with the respective target $>0.2$ and $<-0.2$; almost all correlations are indeed in $[-0.2,0.2]$, and no correlation was outside of $[-0.4,0.4]$. %
}%
\label{fig:models:importances}%
\end{figure}

\section{Predicting whether or not an npm package is runnable}
To answer \RqTwoText~we first need to clarify what we mean by runnable that can be expressed by our features defined in Table~\ref{tab:features} in our experiments.
To do so, we apply the DUO principle (Data Mining Algorithms Using/Used-by Optimizers~\cite{Agrawal2020duo}) to mine insights by using optimisers: %
we employ auto-sklearn~\cite{Feurer2020autosklearn2} to automatically search over the space of machine learning models to achieve an optimised insight, i.e., to achieve a good approximation of the truth while reducing potential limitations of modelling technology and its default or manual configuration. auto-sklearn is a combination of the machine learning library sklearn~\cite{scikit-learn}, the algorithm configurator smac~\cite{Hutter2011smac}, and a set of preprocessing strategies for data.

We investigate three different binary target features as our interpretation of code being runnable, %
focusing on essential aspects of software engineering: %
\begin{enumerate}
\item ableToInstall: while this appears to be a weak condition, it can be a necessary condition for subsequent code development;
\item ableToBuild: stronger than the previous one, and can indicate higher usefulness;
\item ableToExecuteATest: if true, then this can indicate correctness with respect to the specification;
\end{enumerate}

We consider all 104,364 repositories, and we consider all features as inputs except for 
the three ableTo* (to avoid the accidental explanation through strong correlations), as well as the two string-based features; this leaves us with 25 input features. 
For the model search, we perform 5-fold cross-validation. For each fold, auto-sklearn has 10 minutes allocated (single CPU core) to find a well-performing model for a target feature.\footnote{As a performance reference: a random forest from sklearn is typically trained in about one second on our standard laptop.} 
The average F1-scores and accuracy scores over the five folds (per target) are as follows: %
\begin{enumerate}
\item ableToInstall: F1=0.96, accuracy=0.92. This is not too surprising, as 92\% of the data set is installable.
\item ableToBuild: F1=0.83, accuracy=0.73. 72\% can be built. 
\item ableToExecuteATest: F1=0.09, accuracy=0.95. The F1 score is very low, although this is partly due to the imbalanced dataset: it is possible to execute at least one of the tests only for 5\% of the repositories.
\end{enumerate}

To dig deeper into the results, we take (for each target) one of the generated models and investigate the permutation importance of the features~\cite{Breiman2001randomforests}. In this post-hoc approach, a single column of the validation data is randomly shuffled, leaving the target and all other columns in place. This metric measures the effect on the accuracy of predictions in that shuffled data. Due to randomised effects, this process is repeated five times for each feature (sklearn default). %

Figure~\ref{fig:models:importances} shows the results, where larger values indicate higher importance in these tuned models. 
As we can observe, to predict ableToInstall, a large number of features is necessary, with number-based and size-based features being the four most important ones. 
To predict ableToBuild, these features remain important, although the now most important one is hasASourceDirectory. 
To predict ableToExecuteATest, the previously mentioned features remain important, but they are now joined by linesInReadmemd.

Figure~\ref{fig:models:importances} also indicates positive and negative correlations with the respective target features. In general, all correlations with the targets are very weak, i.e., almost all correlations with the targets are within $[-0.2,0.2]$ and are neither marked with a $+$ or a $-$. ableToBuild stands out a bit as a number of features are weakly negatively correlated with it.%

\begin{tcolorbox}
    \textbf{Summary:} Predicting the runnability of a package is viable (i.e., high F1 score). 
    Repository features are particularly important for predicting the runnability.
\end{tcolorbox}

\section{Discussion}
We now discuss our results and provide suggestions for both researchers and practitioners.

\begin{enumerate}

    \item {\textit{Audience perspective matters.}}
    As shown by our analysis of package features, not all features are relevant from all perspectives, and some are even negatively correlated. 
    For example, the user perspective is more interested in documentation than the contributor perspective, which may be interested in workflow characteristics such as the pull-request and issue management systems.
    Our survey results also confirmed that package users are more focused on documentation features.
    For example, one survey respondent mentioned that they adopt a package due to ``\textit{Usefulness, good documentation, recent releases}''.
    On the other hand, contributors are more focused on the contribution guidelines as mentioned by another respondent: ``\textit{simple rules for issues and pull requests}''.

    For practitioners, we suggest that taking into account both the user and contributor may help the overall attractiveness of their Open Source projects. Furthermore, opportunities for future work may be tool support for recommending projects based on the identified features and the different scenarios of these perspectives. Our results confirm that automatically predicting the runnability of a package may be viable.
    For researchers, we suggest that our perspectives open up different scenarios for the different practitioners and a re-evaluation of existing metrics as well as investigating new metrics that can capture these two perspectives.
    The simple heuristics of using the popularity, dependency usage, stars and downloads would need to be re-evaluated, especially for selecting representative samples for empirical studies.

    \item {\textit{Maintaining the runnability of npm packages is non-trivial.}}
    According to our results, the definition of successfully running a project is not trivial.
    Although some work has looked at a single definition, no work had looked at multiple types of runnability.
    For practitioners, we suggest considering these different types when developing projects. 
    This could be an issue if the project is constantly evolving, causing especially code snippets in the documentation (e.g., README) to be outdated. 
    Furthermore, these snippets may be the usage examples or installation instructions, and therefore important for users and contributors.
    Another answer from a survey respondent supports this: ``\textit{One thing that indicates a good package is an example of someone using it to solve an existing problem. For npm packages, those are usually not found in the package's documentation, but on someone's blog}''.

\end{enumerate}

\section{Threats to Validity}

Threats to \textit{construct validity} exist in the appropriateness of our feature list and perspectives.
We mitigated these threats by conducting the survey to verify the relevance of features and perspectives for assessing the quality of packages.
Our feature list is taken from related works related to npm in particular or software reuse in general; however, only a few features have been used for assessing the package quality.
In this case, we received 33 responses to confirm the relevant level of features (i.e., 23 out of 30 features vote agree more than disagree).

Threats to \textit{internal validity} involve the correctness of tools and techniques used in this study.
We use multiple features extracted from \npmDE.
The threat is that sometimes some packages do not have some features or are unavailable to extract, so we applied a filter to remove packages that have at least one missing feature for \RqOne, thus making this threat minimal.

Threats to \textit{external validity} correspond to our ability to generalise results.
Because our data and conclusions are based on a large number of npm packages hosted on GitHub, we cannot generalise our results to all projects in general. Not all npm packages are hosted on GitHub, and some Node.js libraries use alternative package managers like Yarn. Our research is also focused on Node.js and npm only; there are similar package management systems for other languages, such as PyPI for Python and Maven for Java.

\section{Related Work}

\label{sec:related}

\textit{Analysis of Repositories}.
Recent studies have explored dependency networks in different aspects.
Some studies investigated the structure and evolution of the dependency networks and revealed their issues such as dependency hell and technical lag~\citep{Decan:2017, Kikas:2017, Zerouali:ICSR2018}.
Several studies focused on how to detect known security vulnerabilities from third-party libraries in the applications~\citep{Bodin:ASE2020, Zapata:ICSME2018}.
\citet{Cogo:TSE2019} performed an empirical study of dependency downgrades and found that downgrades occur because developers want to avoid some defects from a specific version and some incompatibility issues.
\citet{Hejderup:2018} extracted the call graph for software to build a fine-grained representation of the dependency network.

Social coding in the ecosystem is also an emerging research area. 
Some studies focused on how developers have social interactions with each other~\citep{Constantinou:ISSE2017, Constantinou:SANER2017, Souza:2016, Palyart:2018}.
The relationship between different groups of developers in coding collaboration is explored in various ecosystems~\citep{Palyart:2018, Guercio:2018, Estefo2019}.
\citet{Qiu:2019} focused on how social coding impacts the chances of long-term engagement.
Work such as \citet{ecosystemsIST19} looked at the reference coupling between projects in the ecosystem.
Other studies have shown that the maintenance of these dependencies is critical to the ecosystem \citep{ Zapata:ICSME2018, Decan:2018, Kula:2017, Mujahid:MSR2020}.

\textit{Code Snippet Executability}. 
Documentation is an important part of choosing a library~\cite{Larios:2020} and the popularity of GitHub repositories~\cite{aggarwal2014co}, and example code is in itself an important aspect of good documentation.
Code snippets are primarily used online in documentation, tutorials, and collaborative sites like Stack Overflow to demonstrate how software and APIs should be used; however, not all code snippets are usable as-is~\cite{gistable, docable, YangStackOverflow, NLP2TestableCode}. 
For developers learning APIs, insufficient or inadequate examples can be a major obstacle~\cite{robillard2009makes}, and many code snippets online are incomplete, contain errors, or simply do not work. 
In many cases, code snippets become outdated as software evolves, and despite being the first resource many developers will see, official software documentation is frequently out-of-date, and changes to the software are not immediately reflected in documentation~\cite{lethbridge2003software}. 
There is also little support for developers writing documentation, which makes maintaining up-to-date, executable code snippets in GitHub repositories a non-trivial task~\cite{dagenais2010creating}.

Studies of the executability of online code snippets show that most are not executable; a study of online coding tutorials found that only 26\% of code snippets could be executed successfully, and no tutorial could be executed to completion~\cite{docable}. Gistable~\cite{gistable}, a framework for running Python code snippets found on GitHub using the gist system, found that only 25\% of code snippets were executable by default. These numbers appear to be consistent with other research into code snippet executability~\cite{YangStackOverflow, pimentel2019large, hossain2019executability}, with some variance depending on the language.

\section{Conclusion and Future Directions}
 With more than 1.7 million packages to choose from, selecting a suitable npm package is a difficult task -- and the factors that influence this decision vary based on the developer intentions. Most existing research into library selection has focused on the perspective of a user of a package; in this paper, we have investigated the perspective of user and contributor. 

In this study, we first conducted a survey with 33 respondents to understand which npm package features do practitioners find relevant for assessing package quality.
We found that users and contributors of npm packages share similar views when choosing a package, as half of the considered features belong to both perspectives.
However, the user perspective is more focused on the documentation and the usage of the package.
The contributor perspective, on the other hand, is more focused on contribution guidelines.
Interestingly, developers believe that most repository features do not belong to any perspectives and do not reflect the quality of packages.

We then created the \npmDE~dataset, a dataset containing a curated snapshot of npm packages that the research community can utilise.
We identified a set of 30 features important to one or both perspectives and identified correlations between different perspectives. We found that features from different perspectives are not necessarily correlated; in fact, in some cases, negatively correlated.
This suggests that our different perspectives are important and that trade-offs exist; users and contributors have different priorities. 
This also suggests a need for new metrics to capture these perspectives, which complement the existing features such as popularity, dependency usage, stars, and download count.

We also evaluated the runnability of packages in terms of the runnability of test cases and example code snippets.
Out of 11,127 npm packages that were updated in 2020, we find that 8,199 (73.68\%) are able to be built, and only 607 (7.40\%) are able to pass all test cases.
Out of 104,364 npm packages from our dataset, we found that 97,006 (92.95\%) could be successfully installed, 64,280 (61.59\%) have at least one code snippet in their README, and out of 220,324 code snippets, 33,484 (15.20\%) were able to execute successfully. 
Using this data, we investigated if we could predict the runnability of a package. 
We find that predicting a package runnability depended heavily on what metric of runnability we use; for example, we find that there is a strong correlation between the ability to install a package and its number of files and commits; however, these features are less important in predicting the ability to build a package.%

Our work lays the groundwork for future work on understanding how users and contributors select appropriate npm packages.
We suggest that practitioners should take package audiences into account to help the attractiveness of their packages.
Package owners and contributors should maintain the runnability of their package for attracting and helping new users and newcomer contributors since this is not trivial. 
Potential future avenues for researchers include (1) a package recommendation system based on the runnability of packages and (2) an exploration of new metrics that can measure the package quality with the consideration of both user and contributor perspectives.

\section*{Acknowledgment}

This work was supported by the Japan Society for Promotion of Science (JSPS) KAKENHI Grant Numbers JP18H04094, JP18H03221, 20K19774, 20H05706.

\bibliographystyle{abbrvnat}
\bibliography{bibliography.bib}
\end{document}